\documentclass[twocolumn,aps,showpacs,pra]{revtex4}
\usepackage{amscd}
\usepackage{amssymb}
\usepackage{mathrsfs}
\usepackage{amsfonts}
\usepackage{graphicx}%
\usepackage{dcolumn}
\usepackage{amsmath}

\newcommand{\be}{\begin{equation}}
\newcommand{\ee}{\end{equation}}
\newcommand{\bey}{\begin{eqnarray}}
\newcommand{\eey}{\end{eqnarray}}
\newcommand{\ba}{\begin{array}}
\newcommand{\ea}{\end{array}}
\newcommand{\bi}{\begin{itemize}}
\newcommand{\ei}{\end{itemize}}
\newcommand{\bem}{\begin{enumerate}}
\newcommand{\eem}{\end{enumerate}}
\newcommand{\bw}{\begin{widetext}}
\newcommand{\ew}{\end{widetext}}
\newcommand{\ra}{\rangle}
\newcommand{\la}{\langle}
\newcommand{\pp}{\partial}
\newcommand{\ov}{\overline}

\newcommand{\ww}{\widetilde}

\newcommand{\E}{{\cal E}}

\newcommand{\HH}{{\mathscr{H}}}

\newcommand{\PP}{{\mathcal{P}}}
\newcommand{\D}{{\mathcal{D}}}
\newcommand{\T}{{\mathcal{T}}}
\newcommand{\Pm}{{\mathcal{P}_{\mu}}}
\newcommand{\Pnn}{{\mathcal{P}_{\nu}}}

\newcommand{\R}{{\mathcal{R}}}

\newcommand{\cs}{\mathcal{S}}

\newcommand{\im}{{\rm i}}
\newcommand{\tr}{{\rm Tr}}

\begin{document}
%\draft

 \title{Quantum theory for a total system including one internal measuring apparatus
 }

\author{Wen-ge Wang}
\affiliation{
 Department of Modern Physics, University of Science and Technology of China,
 Hefei 230026, China
 }

 \date{\today}

 \begin{abstract}

 In this paper, we extend the standard formalism of quantum mechanics to a quantum theory
 for a total system including one internal measuring apparatus.
 The internality of the measuring apparatus implies that different decomposition of a given 
 density operator for the internal measuring apparatus 
 into mixture of pure states may have  different physical implications.
 We use `specified mixed-state description' to call a density operator 
 with a specified decomposition into mixture of pure states.  
 The proposed theory has three basic assumptions, which roughly speaking have the following contents:
 (i) Physical states of the total system can be associated with vectors in the total Hilbert space;
 (ii) the dynamical evolution of a state vector obeys Schr\"{o}dinger equation;
 and (iii) under a principle of compatible description and certain non-transition condition,
 a pure-vector description of the total system may imply the existence of certain specified mixed-state
 description.
 The principle of compatible description states that different mathematical descriptions
 for the same physical state of the total system must give
 consistent predictions for results of measurements performed by the internal measuring apparatus.
 This principle imposes a restriction to vectors in the Hilbert space
 and this may effectively break the time-reversal symmetry of Schr\"{o}dinger equation.
% As applications of the proposed theory, the possibility for an $n$-level system to possess
% definite properties is discussed and a master equation is derived for the time evolution of a
% system that may have definite properties for most of the times.

 \end{abstract}

 \maketitle

\tableofcontents

 \section{Introduction}

 \subsection{Motivations}

 The standard formalism of quantum mechanics \cite{Neumann-qm},
 which has passed all experimental tests ever performed,
 is basically a theory for an external observer;
 it gives predictions for measurements performed by an external measuring apparatus.
 A topic that has received extensive attention since the establishment of the formalism,
 with lots of controversy, is the possibility of extending it to a quantum
 theory for an isolated, total system (like the universe) described by an internal observer.
 To achieve this goal, the major difficulty comes from treatment
 of the measuring apparatus, which is now a part of the total system.
 This difficulty is related to the so-called measurement problem, concerning
 the relationship between Schr\"{o}dinger evolution and definite outcomes of measurements.

 The above mentioned problem is of interest not only for pure theoretical reasons,
 but also for a practical reason, concerning designation of small measuring apparatus.
 In recent years, significant progresses have been achieved in technology,
 such that it is now a commonplace
 in labs to observe small systems at the mesoscopic scale, even at the microscopic scale.
 In principle, it is possible to design a measuring apparatus, whose essential part is of
 the microscopic scale.
 A challenging task is to know the condition under which
 a small quantum system may possess some definite properties.
 Our intuition obtained in the macroscopic world is not so helpful for this purpose.

 Lots of efforts have been seen in the attempt of solving the above mentioned problem
 (see, e.g., reviews given in Refs.~\cite{Laloe01,JZKGKS03,Schloss04}),
 most under the name of interpretation of quantum mechanics,
 for example, various versions of Everett's relative-state interpretation (RSI)
 \cite{Everett57,DG73,Deutsch85},
 consistent-histories interpretations (CHI)
 \cite{Griffiths84,Grif96,Griff02,Omnes88,Omnes92,Omnes99,GH90,GH93,Hohen09})
 first proposed by Griffiths \cite{Griffiths84},
 De Broglie's pilot wave theory \cite{deB30} and Bohmian mechanics \cite{Bohm52,BohmB66},
 and dynamical-reduction models \cite{GRW85,GRW86,Pearle89,GPR90,AB09,BG03}.

 There existing so many theories, what is the reason for us to intend to develop another one?
 Besides the fact that there is no commonly-accepted solution to the measurement problem yet,
 one major motivation is that the peculiarity of the internality of the measuring apparatus has not 
 been fully revealed, which should be a key point in solving the measurement problem.

 Another major motivation is as follows.
 Since Schr\"{o}dinger equation has passed all experimental tests ever performed,
 it is reasonable to take this equation as a basic dynamic law \cite{foot-drb}, 
 as done in RSI and effectively so in CHI \cite{Hohen09}.
 Although both RSI and CHI supply quite general frameworks for
 quantum descriptions, neither of them gives a concrete condition, under which
 a considered subsystem may have a definite property (see Appendix \ref{sect-RC-con}).
 To find such a concrete condition is a main motivation of this paper.

 For the simplicity in discussion, in this paper, we focus on the case that the total
 system has one internal observer only \cite{foot-two-R}.
 Since measuring apparatuses that can be controlled by one observer can
 always be regarded as forming a big measuring apparatus, without the loss of generality,
 we assume that there exists only one internal measuring apparatus.
 Further, we assume that one does not need to give the internal observer a special position
 at the fundamental level of the theory,
 thus, the internal observer may be regarded as a part of the environment of the internal measuring 
 apparatus.

% To summarize, the purpose of this paper is to, based on analysis in and understanding of properties of 
% internal measuring apparatus, propose a quantum theory for a total system that includes only 
% one internal measuring apparatus, such that a concrete condition can be given under which the internal 
% measuring apparatus may have a definite property that serves as a measurement outcome. 

 \subsection{A clue suggested by the internality of the measuring apparatus}
 \label{sect-clue}
 
 There exists a basic rule in physics, namely, 
 two mathematical descriptions for a physical system can be regarded as describing the same 
 physical state of the system, if they, as well as their time evolutions determined by the dynamical law, 
 always give compatible predictions for all measurable quantities. 
 This rule implies some significant difference between a quantum theory for a total system including
 a unique measuring apparatus and the usual quantum mechanics for an external observer.
 As to be discussed below, 
 this is closely related to the difference in measurable quantities considered in the two theories.

 In the usual quantum mechanics, 
 measurable quantities are the expectation values of observables of measured systems. 
 Here, as well known, different decompositions of a given density operator for a measured system
 into mixtures of pure states give the same predictions for the expectation values of observables. 
 According to the rule mentioned above, these mixtures of pure states 
 describe the same physical state of the system.
 Thus, a density operator has an unambiguous physical meaning, usually called 
 a mixed state of the measured system.

 On the other hand, for a total system including a unique internal measuring apparatus,  
 the basic measurable quantities are given by definite properties
 of the measuring apparatus, which can be recorded as measurement outcomes. 
 This feature of measurable quantity leads to two properties of the theory, 
 which are significantly different from those in the usual quantum mechanics. 
 First, as to be discussed in detail in Sec.\ref{sect-impli},
 this implies that a density operator for the measuring apparatus does not have an unambiguous physical
 meaning, because different ways of its decomposition into mixtures of pure states may give 
 different predictions for measurement outcomes.
 Therefore, to have a clear physical meaning, the way of its decomposition into mixture of 
 pure states should be specified, which we call a specified mixed-state description.

 Second, as to be shown in detail in Secs.\ref{sect-third-assump} and \ref{sect-TE}, 
 due to the internality 
 and the uniqueness of the measuring apparatus, when appropriate conditions are satisfied, 
 it is possible for a same physical state of the total system to have at the same time 
 both a pure-vector description and a specified mixed-state description. 
 This suggests a way of solving the measurement problem, by relating a pure-vector 
 description of the total system to a specified mixed-state description, where
 each pure state in the mixture predicts certain definite property of the measuring apparatus.

 \subsection{Structure of the paper}
 \label{sect-strategy} 

 \begin{figure}[!t]
  \includegraphics[width=9cm]{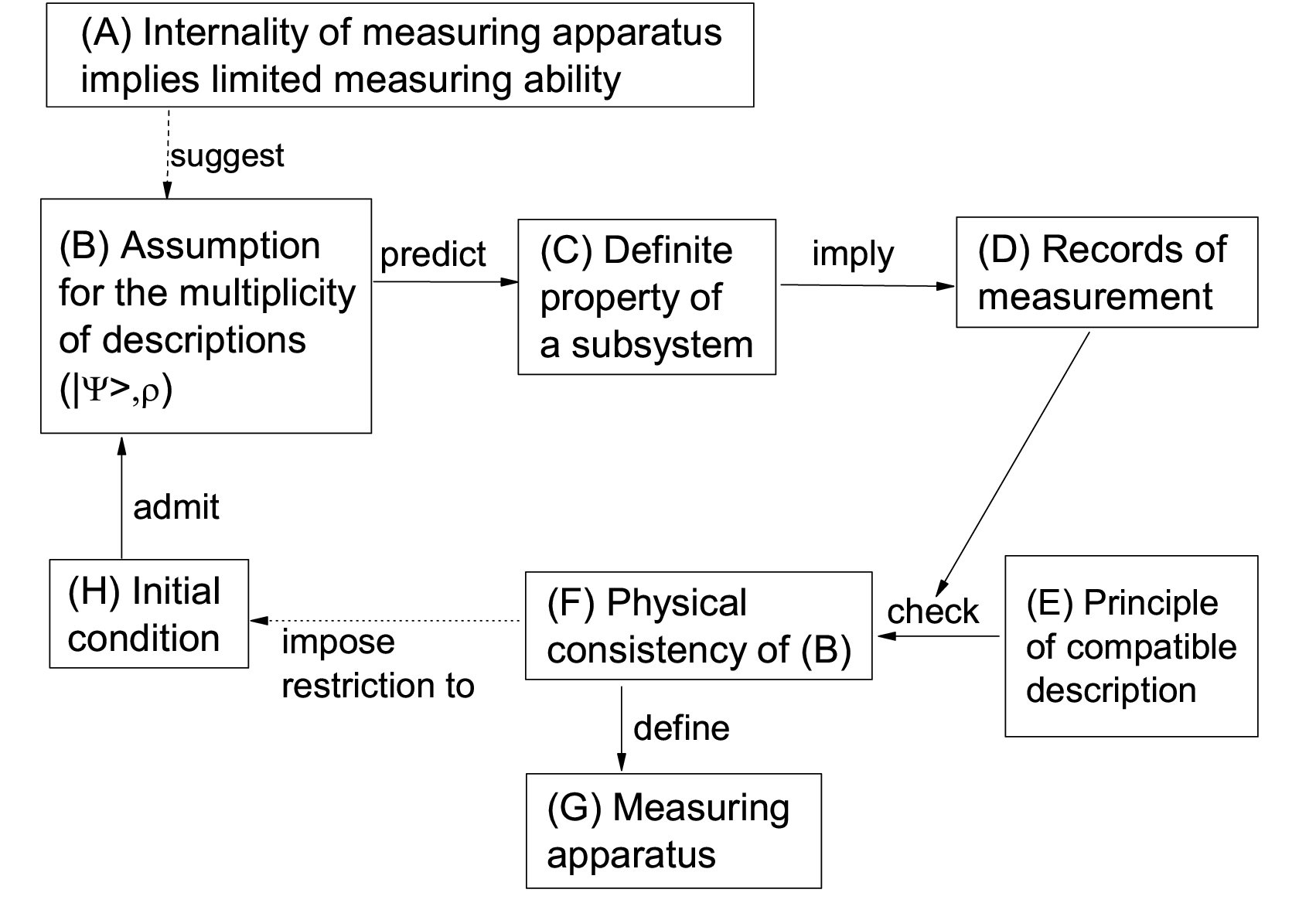}
   \vspace{0.0cm}
 \caption{Schematic plot for a strategy of solving the measurement problem for a total system including
 one internal measuring apparatus.
 It starts from the internality of the measuring apparatus (A), 
 which opens the possibility for a physical state of the total system
 to have both a pure-vector description and a specified mixed-state description at the same time (B).
 The specified mixed-state descriptions may be used to predict some definite properties of
 some subsystem (C), which may give measurement records (D).
 With the help of the predicted measurement records, one may check the physical consistency
 of the multi-descriptions given by the assumption in part B (F),
 by making use of the principle of compatible description (E).
 In the case that the consistency-checking is passed, the considered subsystem may be used as an
 essential part of a measuring apparatus (G).
 This imposes a restriction to the initial condition (H), i.e., the assumption in part B is applicable
 only for those initial conditions that may pass the consistency-checking.
 } \label{fig-loop}
 \end{figure}

 In Sec.\ref{sect-assump}, following in the usual quantum mechanics, 
 we give the first two basic assumptions in the proposed theory,
 namely, the Hilbert space as the state space and Schr\"{o}dinger equation as the dynamic law.
 In Sec.\ref{sect-basis}, we analyze properties of the internal measuring apparatus. 
 The third basic assumption is introduced in Sec.\ref{sect-third-assump}, 
 making use of results given in Sec.\ref{sect-basis}.
 In doing this, we introduce a principle of compatible description,
 stating that two mathematical descriptions for the same physical state of the total system must
 give consistent predictions for measurement results.
 Basically, the third basic assumption states that, subject to the principle of compatible description, 
 when certain condition is satisfied, a pure-vector description of the total system may imply the 
 existence of certain specified mixed-state description. 
 In Sec.~\ref{sect-TE}, we discuss a branching picture of time evolution,
 which is implied by the third basic assumption,
 and derive a mathematical expression for the principle of compatible description.
% This principle in fact imposes a restriction to the initial condition, for which a selected 
% subsystem can be employed as an internal measuring apparatus. 

 The above discussed strategy of developing the theory is schematically plotted in Fig.~\ref{fig-loop}.
 If an initial condition can not pass the consistency-checking given by the principle of compatible
 description, then, this initial condition can not describe a physical state of the total 
 system with a chosen subsystem taken as the internal measuring apparatus. 
% While, it might happen that,  when an other subsystem is taken as the internal measuring apparatus, 
% this initial condition may pass the consistency-checking. 

 Then, we use the proposed theory to discuss various topics.
 In Sec.\ref{sect-meas-axiom}, we discuss a general measurement process and show 
 that POVM measurements can be obtained when measurement schemes are appropriately designed. 
 Sections \ref{sect-isolatable} and \ref{sect-irrever} are devoted to
 some applications of the proposed theory.
 In Sec.~\ref{sect-isolatable}, we show that it is relatively easy for an isolatable system (for example,
 the center-of-mass degrees of freedom of a system) to keep coherence.
% In Sec.\ref{sect-model}, we discuss an $n$-level system and show that it may have definite properties
% under adequate conditions.
 Another application is given in Sec.\ref{sect-irrever}, showing irreversible features of some processes.
 The irreversibility comes from the restriction imposed by the principle of compatible description
 to the initial condition,
 which may effectively break the time-reversal symmetry of Schr\"{o}dinger equation for some processes.
% In Sec.\ref{sect-master}, making use of the third assumption, a master equation is derived
% for the probabilities for an internal measuring apparatus to take definite values.
 Finally, discussions and conclusions are given in Sec.\ref{sect-conclusion}.
 In particular, we discuss the main similarities and differences between the proposed theory
 and the CHI and the many-worlds interpretations of quantum mechanics.

 \section{The first and second basic assumptions}
 \label{sect-assump}

 In this paper, we consider an isolated, total system, which is composed of a system $\R$
 and its environment denoted by $\E$, where the system $\R$ is to be used as the unique measuring apparatus.
 As mentioned in the section of introduction,  we follow the usual quantum mechanics for the first
 two basic assumptions.
 The first one is about the state space.
 \bi
 \item \textbf{Postulate of Hilbert space (HS)}:
 Each physical state of an isolated system, which is described by an internal observer,
 can be associated with a vector in the total Hilbert space $\HH$.
 \ei
 We use $\HH_\R$, $\HH_\E$, and $\HH$ to denote the Hilbert spaces corresponding to the system $\R$,
 its environment $\E$, and the total system $\R+\E$, respectively,
 with $\HH = \HH_{\R}\otimes \HH_{\E}$.
% A density operator is given the ordinary ensemble interpretation.
 We remark that extension of the Hilbert space considered in the usual
 quantum mechanics, which is for descriptions given by an external observer,
 to the total Hilbert space $\HH$ is a non-trivial extension.

 To avoid some ambiguity and difficulty met in a Hilbert space 
 with infinite dimension, in this paper, we consider Hilbert spaces with finite dimensions. 
 But, there is no restriction to the dimensions of the considered Hilbert spaces
 and our discussions will not rely on the exact values of their dimensions. 
 This implies that these dimensions can be as large as one would like them to be,
 hence, discussions to be given below are also valid when the dimensions approach infinity.

 The second basic assumption is about the dynamical law.
 \bi
 \item \textbf{Postulate of Schr\"{o}dinger equation (SE)}:
 The time evolution of a vector description $|\Psi(t)\ra$ of a physical state of the total system
 $\R+\E$ obeys Schr\"{o}dinger equation,
 \be \im \hbar \frac{\pp}{\pp t}|\Psi(t)\ra = H |\Psi(t)\ra . \label{SE}  \ee
 \ei
 Here, we use $H$ to denote the Hamiltonian of the total system $\R+\E$,
 \be H=H_{\R} + H_{\E} + H_I, \ee
 where $H_{\R}$ and $H_{\E}$ are the Hamiltonians of $\R$ and $\E$, respectively,
 and $H_I$ indicates the interaction between $\R$ and $\E$.
 We use $U(t,t_0)$ to denote the unitary evolution operator,
 \be |\Psi(t)\ra = U(t,t_0) |\Psi(t_0)\ra . \label{U} \ee
 In the case of a time-independent Hamiltonian, $U(t,t_0)= e^{-\im H(t-t_0)/\hbar}$.
% As a consequence of Postulate of SE, the time evolution of a density operator $\rho$
% is given by von Neumann equation
% \be \rho (t) = U(t,t_0)\rho(t_0) U^\dag (t,t_0).  \ee

 \section{R-observables of internal measuring apparatus}
 \label{sect-basis}

 In this section, we give analysis in properties of the internal measuring apparatus $\R$,
 which will be useful when introducing the third basic assumption in the next section.

 \subsection{Definite property of measuring apparatus expressed in the Hilbert space}
 \label{sect-internlty-MA}
% \section{Preliminary analysis and structure of the paper}
% \label{sect-PA}

 According to our experiences obtained in labs, the main feature of a measuring apparatus 
 is that, when certain condition is satisfied, \emph{the apparatus may possess a definite property}, 
 which can be recorded as a measurement outcome. 
 The recordability of the definite property implies that it
 can be labeled by a quantity taking discrete values, which we denote by $\mu$ in what follows.
 Indeed, one can never record a continuously-varying quantity.

 For example, the discrete value could be the digital number that appears on a screen of an apparatus
 as the output of measurement,  or the number of ticks of Geiger counters. 
 More subtle is the position of a pointer of a measurement apparatus, which in principle may change 
 continuously. 
 The point here is what we can really record is not the exact position of the pointer, 
 but is the mark on the scale closest to the pointer.
 Obviously, the mark takes discrete values only. 
% Taking discrete values of $\mu$ is also in consistence with an assumption made in the previous section,
% namely, we consider Hilbert spaces with finite dimensions. 
% (In the section of conclusion, we'll give further discussion in 
% whether main results of this paper may be influenced, if $\mu$ may take continuously-varying values.)

 To describe mathematically the above-discussed property labeled by $\mu$, 
 the simplest way is to associate it
 with a division of the Hilbert space of $\R$ into orthogonal subspaces \cite{Neumann-qm},
 which we denote by $\HH_{\R\mu}$.
 The corresponding projection operators for the subspaces, denoted by $\Pm$, satisfy
 \be \Pm\PP_{\nu} = \delta_{\mu \nu } \Pm , \ \ \ \  \sum_{\mu} \Pm = I_{\R},  \label{uni-R} \ee
 where $I_{\R}$ is the identity operator in the space $\HH_{\R}$.
 Correspondingly, the total Hilbert space is also divided into a series of subspaces,
 \be \HH_{\mu } \equiv \HH_{\R \mu }\otimes \HH_{\E} . \ee
 Obviously, $\Pm \otimes I_{\E}$ is the projection operator for the subspace $\HH_{\mu}$,
 where $I_{\E}$ is the identity operator in $\HH_{\E}$.
 For brevity, without the risk of confusion, we also use $\Pm$ to indicate $\Pm \otimes I_{\E}$
 in what follows.
 Related to a set of projection operators $\Pm$, we introduce an observable $A_{\{\mu\}}$
 for the system $\R$,
 \be A_{\{\mu\}} = \sum_{\mu} \mu \Pm , \label{A-mu} \ee
 and call it a \emph{R-observable} of the system $\R$.
% (Explicit definition of R-observable will be given in a later section.)

% A measuring apparatus may have more than one definite properties.
% Our experiences show that these definite properties may exist at the same time, 
% therefore, they must be represented by commuting observables of the measuring apparatus.
% That is, R-observables of $\R$ should be commutable. 
% The second influence is related to the limitedness of the measuring ability of the internal 
% measuring apparatus in measuring its own properties,
% specifically, the only experimental information is given by whether the apparatus has a property 
% definitely related to, or not related to some value(s) of $\mu$.  

 \subsection{R-observable}
 \label{sect-R-observable}

 In this section, we discuss properties a R-observable should have, then, 
 give an explicit definition for this concept.
 For this purpose, we employ the following method. 
 That is, in principle, we may imagine the existence of an external observer
 possessing an external measuring apparatus,
 who has no interaction with the composite system $\R+\E$, such that $\R+\E$ is still isolated.
 Temporarily, we assume that some part of the standard formalism of quantum mechanics can be used
 by the imaginary external observer to give predictions for the composite system $\R+\E$.
 For the consistency of the results to be obtained,
 later (in Sec.\ref{sect-cdes}), we'll show that the temporary assumption can be
 derived from the third basic assumption to be proposed, 
 when the imaginary external observer is regarded as a part of a big system which also includes $\R+\E$.

 Specifically, we make the following temporary assumption for the imaginary external
 observer, denoted by  ${\rm A_T}$.
 \bi
 \item ${\rm A_T}$. 
 The imaginary external observer may use the axiom of measurement
 in the standard formalism of quantum mechanics to predict results of measurements for a
 R-observable of $\R$, when the internal observer predicts that the system $\R$
 has a definite property related to this R-observable.
 \ei
 We assume that the imaginary external observer may also use the
 pure-vector description $|\Psi(t)\ra $, like the internal observer, to describe the state of $\R+\E$.
 In principle, the imaginary external observer may communicate with the internal observer, such that 
 they may compare their measurement results, therefore, the two observers must give compatible predictions
 for definite properties of the internal measuring apparatus $\R$.

 In view of the imaginary external observer, all measurable properties of
 the system $\R$ can be computed from the reduced density matrix $\rho^{re}_{\R}(t) =
 \tr_{\E} |\Psi(t)\ra \la \Psi(t) |$.
 Suppose, in view of the internal observer, the system $\R$ has
 a definite value of $\mu$ with certain probability within some time period.
 Then, for the consistency of the descriptions given by the two observers,
% according to the temporary assumption (${\rm A_T}$) given in Sec.~\ref{sect-external},
 $\rho^{re}_{\R}(t)$ must be block-diagonal with respect to the subspaces $\HH_{\R \mu}$,
 namely, $\Pm \rho^{re}_{\R}(t) \Pnn \propto \delta_{\mu \nu}$ for the same time period.
 This property of the reduced density matrix is usually referred to as decoherence
 induced by environment \cite{Zeh70,Zurek81,JZKGKS03,Zurek03,Schloss04}.

 More exactly, the consistency of the two descriptions discussed above
 requires that $\Pm \rho^{re}_{\R}(t) \Pnn \doteq 0$ for $\mu \ne \nu$.
 Here, $A\doteq B$ means $|A-B|\le \epsilon_x$ (or $\| A-B \| \le \epsilon_x$, or the like),
 where $\epsilon_x \ge 0$ is a small quantity such that its difference from zero
 generates no effect that may be tested by experiments.
 Thus, from the viewpoint of experimental test, $A \doteq B$ is effectively
 equivalent to $A=B$ \cite{footnote-small-e}.

 According to our experiences obtained in labs, the existence of a 
 definite property (not its concrete values) of a measuring apparatus 
 has certain type of initial-condition independence. 
 Here, we should be careful due to the time-reversal symmetry of Schr\"{o}dinger equation, 
 which implies that decoherence can not happen within a finite time period for all initial conditions.
 Since increasing of off-diagonal elements of the reduced density matrix is usually related to
 some coherence
 possessed by the initial vector, in order to determine R-observable, we may consider 
 initially-uncorrelated states, which are described by direct products of vectors in the Hilbert spaces of
 $\R$ and $\E$, respectively, namely, $|\Psi(0)\ra =|\psi^{\R}_0\ra |\phi^{\E}_0\ra$
 with $|\psi^{\R}_0\ra \in \HH_{\R}$ and $|\phi^{\E}_0\ra \in \HH_{\E}$ \cite{foot-ips}.
 In fact, this is the type of initial condition often considered in decoherence theory.

 Summarizing the above discussions and using $C_D$ to denote a (to be determined) condition under which 
 a definite property of $\R$ may appear, 
 we propose the following definition for R-observable.
 \bi
 \item
 A R-observable of $\R$ corresponds to an operator $A_{\{\mu \}}$ satisfying the following requirement:
 For all initial vectors of product form, $|\Psi(0)\ra =|\psi^{\R}_0\ra |\phi^{\E}_0\ra$,
 there exists a decoherence time $\tau_d$, such that if the condition $C_D$ is satisfied for
 a time period $\T =[0,T]$ with $T>\tau_d $, then,  {for} $t \in (\tau_d,T]$,
 \be \Pm \rho^{re}_{\R}(t) \Pnn \doteq 0, \ \ \ \forall  \mu \ne \nu . \ \
 \label{re-dia} \ee
 \ei
 The decoherence time $\tau_d$ is usually a function of the values of $\mu$ of relevance, 
 hence, may be written as $\tau_d(\{\mu \})$.

 Some remarks: Here, the decoherence time $\tau_d$ is defined by the requirement
 $\| \Pm \rho^{re}_{\R}(t) \Pnn \| < \epsilon_x$.
 It is not exactly the same as the decoherence time $\tau_d^u$ usually discussed, which is defined
 by a decay to $1/e$ of the initial value.
 It is easy to verify the relation $\tau_d \sim - \tau_d^u \ln \epsilon_x$.
 Furthermore, the requirement of Eq.(\ref{re-dia}) holding for \emph{all} initial product
 vectors implies that the existence of a R-observable is independent of the concrete status of the
 environment under the specified condition.
 In this sense, a R-observable can be regarded as a system's ``own'' property.

 \subsection{Fine and coarse-grained R-observables}
 \label{sect-fine-R-ob}

 First, we show that coarse-graining of a R-observable gives a new R-observable. 
 To define a coarse-graining of a R-observable $A_{\{\mu\}}$,
 we arrange the labelling $\mu$ into groups labelled by $\eta$, such that
 each $\mu$ belongs to one and only one group $\eta$.
 Then, we define coarse-grained projection operators $\PP_\eta$ as
 \be \PP_\eta = \sum_{\mu \in \eta} \Pm, \label{P-eta} \ee
 which gives a coarse-grained observable $A_{\{\eta\}}$,
 \be A_{\{\eta\}} = \sum_{\eta} \eta \PP_{\eta}.
 \ee
% where $f(\eta)$ is a one-to-one mapping of $\eta$.
 It is easy to verify that, if Eq.~(\ref{re-dia}) is satisfied by the projection operators $\Pm$,
 it is also satisfied by the coarse-grained projection operators $\PP_\eta$.
 Hence, the coarse-grained observable $A_{\{\eta \}}$ is also a R-observable.
 If a R-observable is a coarse-graining of another one, we say that the latter is finer than the former.

 Next, we show that from each two R-observables of $\R$, a finer R-observable may be constructed. 
 In the study of R-observables of $\R$, we consider a fixed set of reduced density matrices 
 $\rho^{re}_{\R}(t)$, namely, those for times $t \in (\tau_d,T]$ obtained
 from initial product vectors at $t=0$.
 We use $S_{re}$ to denote this set of reduced density matrices.
% For each division of the Hilbert space $\HH_\R$ into orthogonal subspaces,
% with respect to which all the matrices in $S_{re}$ are block-diagonal, 
% by the definition of R-observable given in the previous section, there is a corresponding R-observable. 
 According to the definition of R-observable, 
 each matrix $\rho$ in the set $S_{re}$ is block-diagonal with respect to $\Pm$ 
 and has the following expression, 
 \begin{equation}\label{rsp}
 \rho \doteq \sum_\mu \Pm \rho \Pm  \ \hspace{0.8cm} \ \forall \ \rho \in S_{re}.
 \end{equation}
 The block-diagonal form of $\rho$ implies that
 \begin{equation}\label{prhop}
   \Pm \rho \doteq \rho \Pm . 
 \end{equation}
 We use $|i\ra $ to denote normalized eigenstates of a matrix $\rho^0 \in S^{re}$, 
 which has a non-degenerate spectrum $r^0_i$, with $\rho^0 |i\ra = r^0_i |i\ra $.
 (When such a matrix does not exist in $S_{re}$, we use $|i\ra $ to denote eigenstates of 
 a linear combination of some matrices in $S^{re}$, which has a non-degenerate spectrum.)
 Equation (\ref{prhop}) for $\rho^0$ shows that $|i\ra $ are also eigenstates of $\Pm$. 
 Therefore, each $\Pm$ can be constructed by some vectors $|i\ra $, namely, for some set 
 of $|i\ra$, denoted by $q_\mu$, 
 \begin{equation}\label{pmi}
 \Pm = \sum_{i \in q_\mu} |i\ra \la i|. 
 \end{equation}

 Suppose there is another R-observable of $\R$, denoted by $A_{\{\xi \}}$.
 Each projection operator $\PP_\xi$ has also an expression like Eq.(\ref{pmi}), but for a set
 $q_\xi$ of $|i\ra $.
 It is easy to see that 
 \begin{equation}\label{mu-eta}
 \Pm \PP_\xi = \sum_{i \in q_\mu \cap q_\xi} |i\ra \la i|. 
 \end{equation}
 Hence, $\Pm \PP_\xi$ are also projection operators.  
 They give a complete set of projection operators, which we denote by $\{ \PP_{\chi} \}$.
 Thus, each $\PP_{\chi}$ is the `overlap' of some projection operators $\Pm$ and $ \PP_\xi $,
 in short, $\PP_{\chi} = \Pm \bigcap \PP_\xi$.
 This implies that each projection operator $\Pm$ is a coarse-graining of some operators $\PP_\chi$
 (including the possibility that it is one of $\PP_\chi$).
 Therefore, the set $\{\Pm\}$ is either $\{\PP_\chi\}$ or its coarse-graining and 
 similar for $\{\PP_\xi\}$, as a result, both $A_{\{\mu \}}$ and $A_{\{\xi \}}$ are either 
 $A_{\{ \chi \}}$ (given by $\{\PP_\chi \}$) or its coarse-graining.

 Now, we show that $A_{\{ \chi \}}$ is also a R-observable. 
 In fact, multiplying Eq.(\ref{prhop}) by $\PP_\xi$ from the right and making use of the same equation for 
 $\PP_\xi$, it is ready to obtain 
 \begin{equation}\label{}
 \Pm \PP_\xi \rho = \rho \Pm \PP_\xi \ \ \ \forall  \rho \in S_{re}.
 \end{equation}
 Hence, $\PP_\chi \rho = \rho \PP_\chi$ for all matrices $\rho \in S_{re}$.
 Using this result, it is straightforward to verify that Eq.(\ref{rsp}) 
 holds for the set $\{\PP_\chi \}$, hence,  $A_{\{ \chi \}}$ is a R-observable of $\R$.

 Finally, it is ready to show the existence of a finest R-observable of $\R$. 
 The above discussions show that, from each two R-observables, 
 a finer R-observable can be constructed, unless one of the two is already a coarse-graining of the other. 
 Since the Hilbert space $\HH_\R$ has a finite dimension, the process of fining must stop at some stage. 
 Therefore, the system $\R$ must have a finest R-observable, such that all other R-observables
 are coarse-graining of the finest one. 
 Below, we use $A_{\{\mu_f\}}$ to denote the finest R-observable of $\R$, 
 with projection operators $\PP_{\mu_f}$.

 \section{The third basic assumption}
 \label{sect-third-assump}

 In this section, we discuss the third basic assumption.
 The introduction of this assumption will be based on some implications of in the uniqueness 
 and the internality of the measuring apparatus, which we discuss in Sec.\ref{sect-impli}.
% Then, based on results obtained, we propose the third basic assumption in the following two subsections.
 We expect that some properties of
 Schr\"{o}dinger evolution, in particular, decoherence, may be useful when determining the condition for 
 the internal measuring apparatus to possess some definite properties.

 \subsection{Two implications of the internality of the measuring apparatus}
 \label{sect-impli}

 In this section, we continue previous discussions given in Sec.\ref{sect-clue}
 in properties of the measuring apparatus $\R$.
%  which are not possessed by measured systems discussed in the usual quantum mechanics. 
 The uniqueness of the internal measuring apparatus $\R$ implies that 
 all measurements are, first of all, measurements performed by the apparatus $\R$ on itself,
 giving definite values of $\mu$.
 Records of these values of $\mu$ constitute the resource of all experimentally obtainable information.

 One important consequence of the uniqueness and the internality of the measuring apparatus
 is the ambiguity in the physical meaning that can be assigned to a density operator for $\R$.
 To see this point, 
 let us first consider a decomposition of a density operator $\rho$ for $\R$ into a mixture
 $\sum p_\mu |\psi_\mu \ra \la \psi_\mu |$, 
 with the interpretation that the measuring apparatus $\R$ lies 
 in a state $|\psi_\mu\ra $ with a probability $p_\mu$, 
 where $|\psi_\mu\ra $ is a normalized vector in the subspace $\HH_{\R_\mu}$.
 This decomposition implies that the apparatus $\R$ has a definite value $\mu$ of its
 R-observable $A_{\{\mu\}}$, hence, a measurement may be performed by taking 
 record of the value of $\mu$.
 Next, we consider another decomposition of the same density operator, 
 $\rho = \sum p_a |\phi_a \ra \la \phi_a |$,
 with the interpretation that $\R$ lies in a state $|\phi_a\ra $ with a probability $p_a$,
 where no state vector $|\phi_a\ra$ lies in one subspace $\HH_{\R \mu }$.
 In this case, the apparatus $\R$ does not have any definite value of $\mu$, hence, 
 no measurement may be performed with respect to the R-observable $A_{\{\mu\}}$. 
 Clearly, these two decompositions of $\rho$ give different prediction for possible outcomes 
 of measurements performed by the measuring apparatus $\R$.

 Therefore, when using a density operator to describe the internal measuring apparatus,
 the way in which it is decomposed into mixture of pure states should be appropriately specified.
 As mentioned previously, we call a density operator with a specified decomposition into mixture 
 of pure states a \emph{specified mixed-state description}. 
 One meets a similar situation for the total system, when the state of the measuring apparatus
 $\R$ is involved.

 Another consequence, even more important, is the possibility for some 
 pure-vector descriptions and some specified mixed-state descriptions of the measuring apparatus $\R$
 to be physically compatible. 
 To illustrate this, let us consider
 a pure-state description $|\psi\ra = |\psi_{\mu_1}\ra + |\psi_{\mu_2}\ra $ for $\R$,
 where $|\psi_{\mu_1}\ra \in \HH_{\R_{\mu_1}}$ and $|\psi_{\mu_2}\ra \in \HH_{\R_{\mu_2}}$,
 and a specified mixed-state description $\rho = |\psi_{\mu_1}\ra \la \psi_{\mu_1} | + 
 |\psi_{\mu_2}\ra \la \psi_{\mu_2} |$.
 The main difference between the two descriptions lies in the coherence between 
 $|\psi_{\mu_1}\ra$ and $|\psi_{\mu_2}\ra$ in the description $|\psi\ra $, while there existing no such 
 coherence in the description $\rho$.

 The point here is that the only way of experimentally testing the coherence between 
 $|\psi_{\mu_1}\ra$ and $|\psi_{\mu_2}\ra$ in $|\psi\ra $ is through measurements performed by $\R$
 on its own definite properties, meanwhile, a measuring apparatus has only limited ability in 
 measuring its own properties. 
 In fact, as known in decoherence theory \cite{Zeh70,Zurek81,JZKGKS03,Zurek03,Schloss04}, 
 as far as only properties of $\R$ are concerned, under appropriate 
 conditions, environmentally-induced decoherence may sufficiently suppress the coherence between 
 $|\psi_{\mu_1}\ra$ and $|\psi_{\mu_2}\ra$ in $|\psi\ra $.
 When this happens, the measuring apparatus $\R$ will be unable 
 to test any effect of the coherence, as a result, 
 it is possible for the two descriptions $|\psi\ra $ and $\rho$ to be physically compatible.

 The above-discussed relation between $|\psi\ra $ and $\rho$ suggests a way
 by which definite properties of the measuring apparatus may be predicted.
 Namely, a 
 specified mixed-state description for the total system may be predicted from a pure-vector description, 
 the latter of which is given by unitary evolution of an initial vector,
 according to the postulate of SE \cite{KB07}.
 In the following two sections, we'll follow this idea to propose the third basic assumption. 
 Moreover, for the consistency of the theory, one should further consider the physical compatibility of 
 the time evolutions of the two descriptions and this will be discussed in detail in Sec.\ref{sect-TE}.

 \subsection{Assumption of specified mixed-state description (MsD)}
 \label{sect-assum-MsD}
% \subsection{Mixed-state description}
 \label{sect-explicit}
 \label{sect-CD}

 Based on discussions given in the previous section, we propose that
 a major part of the third basic assumption has the following contents:
 If certain condition $C_D$, which has used in the definition of R-observable 
 in Sec.\ref{sect-R-observable}, is satisfied by Schr\"{o}dinger evolution $|\Psi(t)\ra $ 
 for a time period $t \in \T \equiv [0,T]$, then, at the time $t=T$,
 the total system has another (specified mixed-state) description that 
 it is described by some vector $|\Psi_\mu(T)\ra \in \HH_\mu$ with some probability $p_\mu$.
 The reason of considering a time period $\T$, but not an instant, is that 
 we expect an important role played by decoherence, the happening of which needs a time period.

 We first determine the expressions of $|\Psi_\mu(T)\ra $ and $p_\mu$. 
 For this purpose, we employ the method of
 considering an imaginary external observer, which has been used in Sec.\ref{sect-R-observable}. 
 According to the temporary assumption ${\rm A_T}$, the imaginary external observer predicts that,
 if a measurement is
 performed on the observable $A_{\{\mu \}} \otimes I_{\E}$, there is a probability
 $\la \Psi(T) | \Pm | \Psi(T)\ra / \la \Psi(T)|\Psi(T)\ra $ for a value $\mu$ to come out,
 meanwhile, the system $\R+\E$ lies in a state described by $\Pm |\Psi(T)\ra $.
 For the consistency between this prediction of the imaginary external observer
 and the prediction of $|\Psi_\mu(T)\ra $ and $p_\mu$ by the internal observer, we have
 \be |\Psi_\mu(T)\ra = \frac{ \Pm |\Psi(T)\ra}{\| \Pm |\Psi(T)\ra \|} , \
 p_\mu = \frac{ \la \Psi(T) | \Pm | \Psi(T)\ra }{ \la \Psi (T) |\Psi(T) \ra } . \label{Psi-rho}
 \ee

 Thus, we reach the following assumption of specified mixed-state description (MsD). 
 \bi
 \item \textbf{Assumption of specified MsD}:
 If the total system $\R+\E$ has a description $|\Psi(t)\ra $ within a time period $\T = [0,T]$, 
 which satisfies a condition $C_D$,  then, at the time $T$ the same physical state of
 the total system has another description that
 it is described by $| \Psi_\mu(T)\ra $ with a probability $p_\mu $ in  Eq.(\ref{Psi-rho}).
 \ei
 That is, at the time $T$, in addition to the Schr\"{o}dinger evolution
 $|\Psi(T)\ra $, the total system also has the following specified mixed-state description,  
 \bey  \rho(T) = \frac 1{\la \Psi(T) |\Psi(T) \ra }  \sum_\mu \Pm |\Psi(T)\ra  \la \Psi(T)| \Pm . \ \
 \label{rho-ori} \eey
% Note that as a specified mixed-state description, here the density matrix $\rho(T)$ has the 
% specific decomposition into mixture of pure states given on the right hand side of Eq.(\ref{rho-ori})

 Now, we discuss the condition $C_D$.
 According to the assumption of specified MsD, it is the condition under which 
 a pure-vector description $|\Psi(t)\ra $ of the total system may imply the existence of the 
 specified mixed-state description $\rho(T)$ in Eq.(\ref{rho-ori}).
 We note that, following arguments similar to those given in Sec.~\ref{sect-R-observable}, 
 Eq.(\ref{re-dia}) must hold for $|\Psi(T)\ra$.
 This requirement can not be fulfilled, if $\| \PP_{\nu} H \Pm |\Psi(t)\ra \|$ of $\nu \ne \mu$
 is not negligibly small before $t=T$.
 In fact, if there had been non-negligible transition among the subspaces $\HH_\mu$ before $t=T$,
 usually the interaction generates non-negligible elements of $\Pm \rho^{re}_{\R}(T) \Pnn $
 for $\mu \ne \nu$.
 Therefore, a necessary part of the condition $C_D$ should be that,
 for certain time period before $t=T$, there is negligible transition among the subspaces $\HH_\mu$.

 Since Eq.~(\ref{re-dia}) in fact represents a decoherence effect and it usually takes a
 decoherence time $\tau_d$ for decoherence to happen,
 the above-discussed time period before $t=T$ should not be shorter than the decoherence time $\tau_d$.
 Therefore, generally, the condition $C_D$ requires negligible transition among the subspaces
 $\HH_\mu$ for a time period $\T = [0,T]$ with $T \ge \tau_d$.
 Writing the above results explicitly, we have
 \be  \frac{1}{\la \Psi |\Psi\ra^{1/2}} \| \PP_{\ov \mu} U(t,0) \Pm
 |\Psi(0)\ra \| \doteq 0, \  \forall \mu , \   t\in \T,  \label{NCC} \ee
 where $\PP_{\ov \mu} \equiv I-\Pm$ and $T \ge \tau_d$.
 We call Eq.~(\ref{NCC}) the \emph{non-transition condition} for $|\Psi(t)\ra $ with respect to
 the R-observable $A_{\{\mu \}}$.
% Note that the non-transition condition also implies stableness of the corresponding property of $\R$.
 We do not see any other element that must be included in the condition $C_D$,
 therefore, we assume that 
 \bi
 \item condition $C_D$ = non-transition condition (\ref{NCC}).
 \ei
 Satisfaction of the non-transition condition implies that
 $\Pm |\Psi(t)\ra \doteq U(t,0)\Pm | \Psi(0)\ra $ for $t\in \T$.

 \subsection{The third basic assumption}
 \label{sect-general-3a}

 As discussed previously, 
 a necessary and sufficient condition for a same physical state of the total system $\R+\E$
 to have two mathematically different descriptions at the same time, is that the two descriptions are
 experimentally compatible with respect to measurement results of the internal measuring apparatus.
 This gives the following principle.
 \bi
 \item  \textbf{The principle of compatible description}:
 Different mathematical descriptions for the same physical state of the total system $\R+\E$
 must give compatible predictions for the probabilities for the system $\R$ to have definite properties.
 \ei
 We call this a principle, because it must be obeyed in all physical theories.

 Now, we are ready to propose the third basic assumption, which completes the basic structure 
 of proposed theory. 
 \bi \item \textbf{The third basic assumption}:
 The assumption of specified MsD is applicable to a state vector of the total system, 
 subject to the principle of compatible description. 
 \ei
 To put it more explicitly, if applications of the assumption of specified MsD to a vector $|\Psi\ra $,
 as well as all related time evolutions, 
 do not lead to confliction with the principle of compatible description,
 then, the assumption of specified MsD is applicable to this vector.

 Below are some remarks and comments.
 (1) The principle of compatible description guarantees the physical consistency of the theory.
% Here, special attention must be paid to this principle, because one physical state of the total system
% may have a pure-vector description and a specified mixed-state description at the same time.

 (2) Most of the contents in the assumption of specified MsD are given based on
 our experiences obtained in labs and an appropriate part of the standard formalism of quantum mechanics.
 The part lacking such a sound basis is the assumption that the non-transition condition
 is sufficient for the condition $C_D$.
 Further discussions about this point will be given in Sec.\ref{sect-NTC}.

 (3) The third basic assumption implies an unusual mathematical 
 structure of the theory, in the sense that it is composed of two involved parts.
 On one hand, applicability of the assumption of specified MsD is subject to satisfaction of 
 the principle of compatible description. 
 On the other hand, to know whether the principle of compatible description is satisfied or not, 
 the assumption of specified MsD must be used to give predictions. 
 This feature has its origin in the two-fold roles played by the internal measuring apparatus,
 namely, it is a part of the described total system, meanwhile, it is the unique system
 that may check the physical consistency of descriptions of the total system.

 \section{Time evolution and restriction in initial condition}
 \label{sect-TE}

 In this section, we discuss properties of time evolution implied by the third basic assumption,
 as well as a restriction in the measuring apparatus and the initial condition. 
 In order to describe physical processes of the total system
 starting from an initial condition $|\Psi(t_0)\ra $, 
 one should first choose a subsystem $\R$, which is expected to be used as the internal measuring apparatus. 
 Then, one may study predictions given by the assumption of specified MsD for the time evolution of 
 this initial condition.
 In the case that the obtained predictions satisfy the principle of compatible description, 
 according to the third basic assumption, 
 the assumption of specified MsD is applicable to this initial condition with the chosen subsystem $\R$
 used as the internal measuring apparatus;
 otherwise, the chosen subsystem $\R$ can not be used as a measuring apparatus under this 
 initial condition.

% Specifically, in this section, we derive an explicit expression of the principle of compatible 
% description.
% This expression imposes a restriction to the initial condition, for which a given subsystem $\R$
% can be used as an internal measuring apparatus. 
% We also give further discussions about the non-transition condition.

 \subsection{Time evolution under the non-transition condition}
 \label{sect-t-nt}

 It would be useful to first give some discussions in the unitary evolution of the total system, 
 when the non-transition condition is satisfied. 
 Suppose the non-transition condition Eq.(\ref{NCC}) is satisfied within a time period $\T=[0,T]$.
 This implies that $ \PP_{\ov \mu} H \Pm |\Psi(t)\ra \doteq 0$, hence,
 \be \label{Pm-HI} \Pm H|\Psi(t)\ra \doteq \Pm H \Pm |\Psi(t)\ra \ \ \ \ \forall \ \mu ,
  \ t \in \T .  \ee
% Note that $\Pm$ is effectively commutable with $H$ for $|\Psi(t)\ra $ within this time period.
 Then, multiplying Schr\"{o}dinger equation (\ref{SE}) by $\Pm$ from the left,
 we get the following equation of motion,
 \be \im \hbar \frac{\pp}{\pp t} |\Psi_\mu(t)\ra \doteq  H_\mu |\Psi_\mu(t)\ra  \ \
 \text{for}  \ t \in \T ,
 \label{SE-Pmu} \ee
 where $|\Psi_\mu (t)\ra = \Pm |\Psi(t)\ra $ and
 \be  H_\mu \equiv \Pm H \Pm  = \Pm H_{\R}\Pm + \Pm H_I \Pm + H_{\E}\Pm, \label{wwHmu}
 \ee
 which is an operator acting in the subspace $\HH_{\R \mu} \otimes \HH_\E$.
 Equation (\ref{SE-Pmu}) has the following formal solution,
 \be |\Psi_\mu(t)\ra \doteq \exp \left ( -\im H_\mu t/\hbar \right ) |\Psi_\mu(0)\ra  \ \
 \text{for}  \ t \in \T .
 \label{solu-mu} \ee

 The non-transition condition (\ref{NCC}) is equivalent to the relation
 $\PP_{\ov \mu}H \Pm |\Psi(t)\ra \doteq 0$, hence, is equivalent to
 \be  \PP_{\ov \mu} H_{\R} \Pm |\Psi(t)\ra \doteq - \PP_{\ov \mu} H_I \Pm |\Psi(t)\ra  \ \ \ \forall
 \mu, \ t \in \T .
 \label{PHPm} \ee
 The operator $\PP_{\ov \mu} H_{\R} \Pm$ on the left hand side of Eq.(\ref{PHPm}) has
 trivial action in the Hilbert space $\HH_{\E}$, while the operator
 $ \PP_{\ov \mu} H_I \Pm$ on the right hand side has non-trivial action in $\HH_{\E}$.
 Hence, generally, for the relation in Eq.(\ref{PHPm}) to hold, its two sides must be effectively
 equal to zero, that is, for all values of $\mu$ and for $t \in \T$,
 \bey \label{PPHS} \PP_{\ov \mu} H_{\R} \Pm  |\Psi(t)\ra \doteq 0,
 \\ \PP_{\ov \mu} H_I \Pm  |\Psi(t)\ra \doteq 0  .
 \label{PPHI} \eey
 An important case, in which Eq.(\ref{PPHS}) is satisfied, is that
 $\PP_{\ov \mu} H_{\R} \Pm \doteq 0$, or equivalently,
 \be [H_\R,  A_{\{\mu \}}] \doteq 0 .  \label{sta-00} \ee
 In this case, the subspaces $\HH_{\R\mu}$ are effectively eigen-subspaces of $H_{\R}$
 and the corresponding definite property of $\R$ is stable as long as the interaction is weak.

 \subsection{Tree structure of branching for time evolution}
 \label{sect-branch}

 In this section, we discuss a tree structure formed by the components of the
 specified mixed-state descriptions predicted by the assumption of specified MsD.
 Let us consider an initial state of the total system, which is described by 
 a normalized vector $|\Psi(t_0)\ra $.
 According to the postulate of SE, for $t>t_0$, the state vector has Schr\"{o}dinger evolution, 
 $|\Psi(t)\ra = U(t,t_0)|\Psi(t_0)\ra $.
 Suppose the non-transition condition (\ref{NCC}) is satisfied for a R-observable $A_{\{\mu_{(1)}\}}$
 within a time interval $[\tau_1,\ww\tau_1]$, with $\ww\tau_1-\tau_1 \ge \tau_d$.
 Then, according the assumption of specified MsD, for the time $t_1 = \tau_1+\tau_d$,
 besides the pure-vector description $|\Psi(t_1)\ra $,
 the total system also has the following specified mixed-state description [see Eq.(\ref{rho-ori})],
 \be \rho(t_1)  =  \sum_{\mu_{(1)}} \PP_{\mu_{(1)}} |\Psi(t_1)\ra \la \Psi(t_1)| \PP_{\mu_{(1)}}^\dag ,
 \label{r-t10} \ee
 namely, with a probability $\la \Psi(t_1) |\PP_{\mu_{(1)}} |\Psi(t_1)\ra $, the total system 
 lies in a state described by $\PP_{\mu_{(1)}} |\Psi(t_1)\ra$ and has a definite value $\mu_{(1)}$.

\begin{figure}[!t]
  \includegraphics[width=\columnwidth]{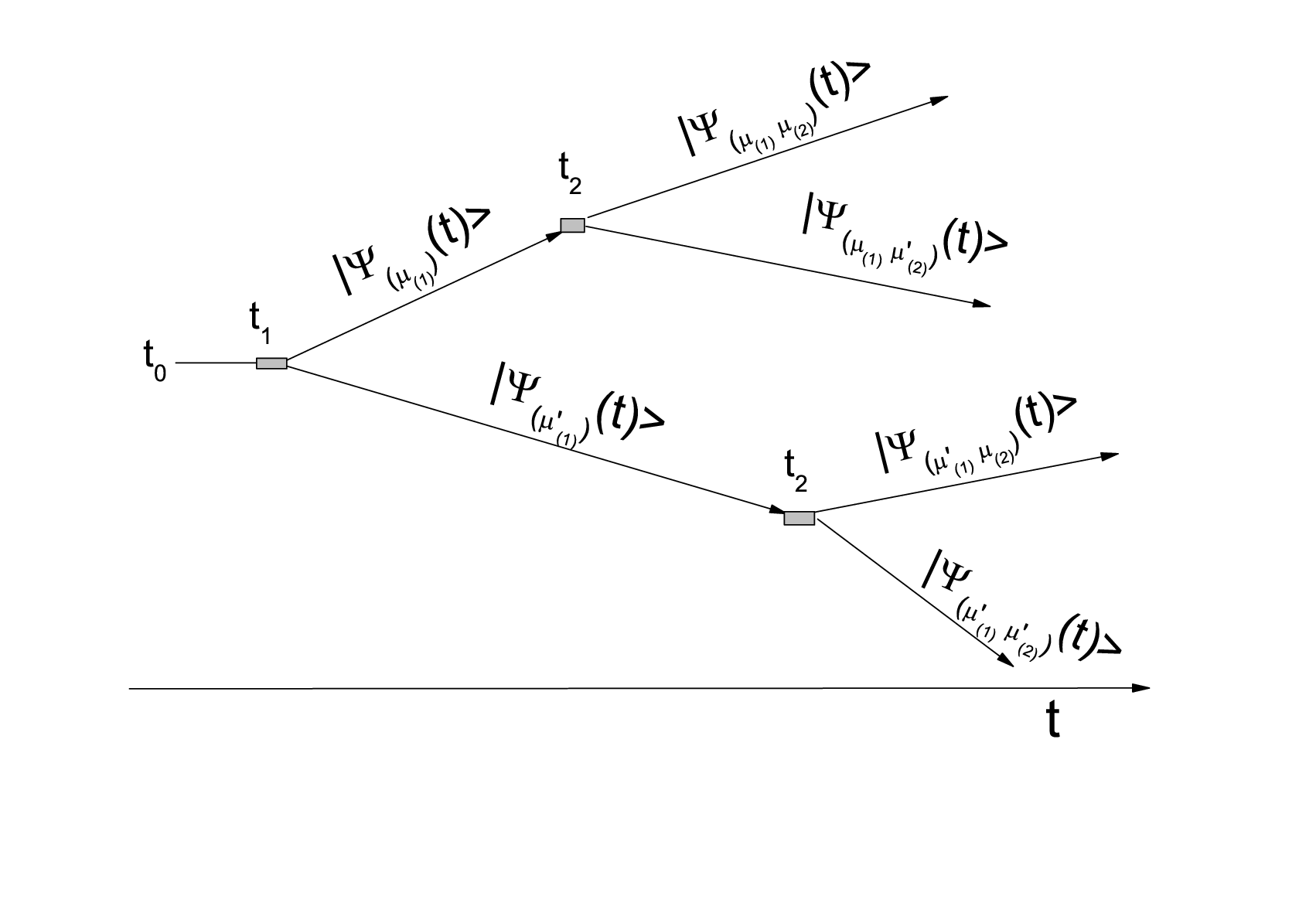}
   \vspace{-0.4cm}
 \caption{Schematic plot for the tree structure of time evolution.
 Each small square represents a splitting point of a component
 and each short line with direction represents Schr\"{o}dinger evolution of a component (branch).
 Specifically, the tree plotted in this figure starts from an initial vector
 $|\Psi(t_0)\ra $, then, evolves unitarily to a time $t_1$ and
 splits into components $|\Psi_{(\mu_{(1)})}(t_1)\ra $.
 Then, after some period of unitary evolution, each component (branch) splits at a time $t_2$,
 giving a sub-branch $|\Psi_{(\mu_{(1)}) (\mu_{(2)}) }(t)\ra $.
 A series of successive branches going from left to right form a path.
 Paths with different values of $\mu_{(1)}$ may have different times $t_2$ and different labeling
 $\mu_{(2)}$.
 } \label{fig-bra}
 \end{figure}

 Each component in the above specified mixed-state description evolves obeying 
 Schr\"{o}dinger equation, hence,
 \be \rho(t)  =  \sum_{\mu_{(1)}} |\Psi_{(\mu_{(1)})}(t)\ra \la\Psi_{(\mu_{(1)})}(t)|,  \ \ \text{for}
 \  t >t_1,
 \label{r-t1} \ee
 where
 \be |\Psi_{(\mu_{(1)})} (t)\ra = U(t,t_1)\PP_{\mu_{(1)}} |\Psi(t_1)\ra  . \label{Psi-mu1}
 \ee
 For brevity, one may say that the vector $|\Psi(t)\ra$ ``\emph{splits}'' into
 the components $|\Psi_{(\mu_{(1)})} (t)\ra$ at $t=t_1$.
 This feature is schematically plotted in Fig.~\ref{fig-bra}.

 Note that, although $|\Psi_{(\mu_{(1)})} (t)\ra \in \HH_{\mu_{(1)}}$ for $t \in (t_1,\ww\tau_1)$,
 beyond the time $\ww\tau_1$, it is not necessary for the vector
 $|\Psi_{(\mu_{(1)})} (t)\ra$ to lie in the subspace $\HH_{\mu_{(1)}}$, since the non-transition
 condition is not satisfied beyond $\ww \tau_1$.
 For this reason, in the subscript of $\Psi $ we write $\mu_{(1)}$ in a pair of parentheses.

 Suppose for a component $|\Psi_{(\mu_{(1)})} (t)\ra$ of $t>\ww\tau_1$,
 the non-transition condition (\ref{NCC}) is satisfied for a R-observable $A_{\{\mu_{(2)}\}}$
 within a time period $[\tau_2,\ww\tau_2]$, with $\ww\tau_2-\tau_2 \ge \tau_d$.
 Then, $|\Psi_{(\mu_{(1)})} (t)\ra$ may split at the time $t_2 = \tau_2+\tau_d$,
 giving the following specified mixed-state description according to the assumption of specified MsD,
 \be \rho(t)  =  \sum_{\mu_{(2)}} |\Psi_{(\mu_{(1)}\mu_{(2)})}(t)\ra
 \la\Psi_{(\mu_{(1)}\mu_{(2)})}(t)|  \ \ \text{for} \  t >t_2,
 \label{r-t0} \ee
 where
 \be |\Psi_{(\mu_{(1)}\mu_{(2)})} (t)\ra = U(t,t_2)\PP_{\mu_{(2)}} |\Psi_{(\mu_{(1)})} (t_2)\ra\ra  .
 \label{Psi-mu12} \ee
 These features are also plotted in Fig.~\ref{fig-bra},
 where components like $|\Psi_{(\mu_{(1)})} (t)\ra$ are indicated by short lines.

 Proceeding with the above procedure,
 with increasing time, splitting of components may happen again and again.
 Since subscripts of $\Psi$ will become even longer,
 for brevity, we use $\alpha$ to indicate a sequence of splittings and call it
 a \emph{path} of splitting.
 Explicitly, we have
 \be \alpha = \left ( \mu_{(1)}^\alpha(t^{\alpha}_1) \to \mu_{(2)}^\alpha(t^{\alpha }_2) \to
 \ldots \to \mu_{(n)}^\alpha(t^{\alpha }_n) \right )
 \label{path-mu} \ee
 for a path with $n$ splittings.
 For example, for $n=2$ we have $|\Psi_\alpha(t) =|\Psi_{(\mu_{(1)}^\alpha \mu_{(2)}^\alpha )}(t)\ra$.
 Along a path $\alpha$, around the $i$-th spitting, the non-transition condition (\ref{NCC}) is
 satisfied for a R-observable $A_{\{\mu_{(i)}^\alpha\}}$
 within a time period $[\tau^{\alpha}_i,{\ww\tau^{\alpha }_i}]$ with ${\ww\tau^{\alpha }_i}-
 \tau^{\alpha}_i \ge \tau_d$; the $i$-th spitting happens at the time $t_i^\alpha = \tau^{\alpha}_i
 +\tau_d$.

 Before continuing our discussion, it is useful to give some remarks about the notations used:
 (1) In the general situation here, a superscript $\alpha$ is added to
 $\mu$, $\tau$, and $t$ belonging to a path $\alpha$, since the values of $\tau_i$, $\mu_{(i)}$, and
 $t_i$ may be different along different paths $\alpha$.
 (2) We write the number $i$ in the subscript of $\mu^\alpha_{(i)}$ within parentheses,
 to indicate explicitly that the corresponding projection operators $\Pm$ at different splitting
 points $i$ may be different.
 (3) A path $\alpha$ is in fact a function of the time $t$, hence, sometimes we write $\alpha(t)$.

% The splitting in $\alpha_n$ happen at the times $t_0,t_1,\ldots ,t_n$
% and generate new components indicated by $m^{(0)}, m^{(1)}, \ldots, m^{(n)}$, respectively.
% A splitting time $t_n$ is a function of the previous path, therefore, it can be
% written as $t_n(\alpha_{n-1})$.

 The short lines and small squares in Fig.~\ref{fig-bra} form a structure like a tree.
 For this reason, we call one set of compatible splitting points and components,
 which stem from the same initial condition, like those shown in Fig.~\ref{fig-bra}, a \emph{tree}
 and denote it by $\Upsilon$.
 We call a component in a tree, represented by a short line in the figure,
 a \emph{branch} of the tree.

 Similar to Eqs.~(\ref{Psi-mu1}) and (\ref{Psi-mu12}),
 we get the following explicit expression for a component at a time $t$,
 reached through a path $\alpha$,
 \bw
 \bey  |\Psi_{\alpha}(t)\ra = U(t,t^{\alpha }_n) \PP_{\mu_{(n)}^\alpha}
 U(t_{n}^\alpha,t_{n-1}^\alpha) \PP_{\mu_{(n-1)}^\alpha}
  \cdots U(t_{i+1}^\alpha,t_{i}^\alpha) \PP_{\mu_{(i)}^\alpha} U(t_{i}^\alpha,t_{i-1}^\alpha)
  \cdots U(t_{2}^\alpha,t_{1}^\alpha) \PP_{\mu_{(1)}^\alpha} U(t_1^\alpha,t_0) |\Psi(t_0)\ra .
 \label{Psi-alpha} \eey
 \ew
 At the time $t$, the state of the total system can be described
 by both the vector $|\Psi(t)\ra =U(t,t_0) |\Psi(t_0)\ra $ and the following specified mixed-state
 description,
 \be \rho_\Upsilon(t)
 %= \sum_{n=0}^{M} \sum_{\alpha_n} |\Psi_{\alpha_n}(t)\ra \la \Psi_{\alpha_n}(t)|
 = \sum_{\alpha \in \Upsilon} |\Psi_{\alpha}(t)\ra \la \Psi_{\alpha}(t)|.
 \label{rho-fi} \ee
 We remind that the specified mixed-state description in Eq.(\ref{rho-fi}) has the specific meaning
 that the total system lies in a state $|\Psi_{\alpha}(t)\ra$, reached through a path $\alpha$, 
 with the probability
 \be P_{\alpha}(t) = \la \Psi_{\alpha}(t) |\Psi_{\alpha}(t)\ra . \label{Pan} \ee
 It is easy to verify that
 \be \sum_\alpha \la \Psi_{\alpha}(t) |\Psi_{\alpha}(t)\ra =1. \label{unit-P} \ee
 Using the relation $ \sum_{\mu_{(i)}^\alpha} \PP_{\mu_{(i)}^\alpha} =I \label{Pmn=1} $
 for each pair $(\alpha , i)$, it is easy to verify
 \be |\Psi(t)\ra = \sum_{\alpha \in \Upsilon } |\Psi_\alpha (t)\ra .  \label{PUP1} \ee

 \subsection{Fine and coarse-grained trees}
 \label{sect-fine-tree}

 A splitting of a branch along a path implies a change in the description
 of the same physical state of the total system, but not a physical change of the total system. 
 In fact, starting from the same initial state $|\Psi(t_0)\ra $, there may exist many trees,
 because the only requirement for a branching to happen is satisfaction of the non-transition condition.
 For example, the $i$-th splitting time $t_i^\alpha$ along a path $\alpha$ may in fact
 take any value between $\tau_i^\alpha +\tau_d$ and $\ww \tau_i^\alpha$;
 meanwhile, if the non-transition condition is satisfied for one R-observable, then,
 it is also satisfied for a coarse-graining of the R-observable.
 Furthermore, since a branching is just a change of description, when the non-transition condition 
 is satisfied, one has the freedom of not choosing the happening of branching.

 The above-discussed multiplicity in the tree description is not so strange as it looks like at the 
 first sight, since there exists coarse-graining relationship among them.
 In fact, among the trees starting from the same initial condition,
 there exists a finest tree, such that other trees are its coarse-graining. 
 This is a requirement of the principle of compatible description.

 Let us use $\Upsilon_f$ to denote a tree description of the total system starting from the initial 
 condition $|\Psi(t_0)\ra $, which is obtained by taking branch-splitting as fine as possible 
 whenever a splitting is possible. 
 We use $\beta_f$ to denote the corresponding paths.
 Below, we show that this tree $\Upsilon_f$ is the finest tree starting from $|\Psi(t_0)\ra $.

 Let us consider an arbitrarily-chosen tree description $\Upsilon$ with paths $\alpha$, 
 which also starts from $|\Psi(t_0)\ra $,
 hence, describes the same physical state of the total system as $\Upsilon_f$ does. 
 Suppose the first branching of $\Upsilon$ happens at a time $t_1$ for a R-observable $A_{\{\mu\}}$.
 Thus, at the time $t_1$, 
% in addition to the pure-vector description $|\Psi(t)\ra $, 
 the total system has the specified mixed-state description that, 
 with the probability $p_\alpha(t_1) = \la \Psi(t_1) |\Pm |\Psi(t_1)\ra $,
 it is described by [see Eq.(\ref{r-t10})]
 \be  |\Psi_\alpha(t_1)\ra =\Pm |\Psi(t_1)\ra , \label{Psiat1} \ee
 possessing a definite value of $\mu$, where $\alpha$ is just $\mu$ in this case.

 Meanwhile, according to the tree $\Upsilon_f$, with the probability $\la \Psi_{\beta_f}(t_1)
 | \Psi_{\beta_f}(t_1)\ra $, the total system is described by $|\Psi_{\beta_f}(t_1)\ra $.
 A component $|\Psi_{\beta_f}(t_1)\ra $ is usually an eigenvector of more than one R-observables
 and we use $A_{\{ \nu^{\beta_f} \}}$ to denote the finest one among these R-observables, 
 with $\nu^{\beta_f}$ denoting the corresponding eigenvalue.

 The point is to note the following requirement of the principle of compatible description. 
 That is, if $\la\Psi_{\beta_f}(t_1)|\Psi_\alpha(t_1)\ra \ne 0$, 
 then, the two definite values $\mu$ and $\nu^{\beta_f}$ must be compatible. 
 In terms of projection operators, this implies that either the projection operator $\PP_{\mu}$ 
 is a coarse-graining of some projection operators $\PP_{\nu^{\beta_f}}$, 
 or the projection operator $\PP_{\nu^{\beta_f}}$ 
 is a coarse-graining of some projection operators $\PP_{\mu}$. 
 Since, by definition, $\Upsilon_f$ is obtained by the finest R-observable whenever possible, 
 we must have the former case. 
 Hence, for each component $|\Psi_{\beta_f}(t_1)\ra$ and an arbitrary value $\mu$,
 \be \Pm|\Psi_{\beta_f}(t_1)\ra = \ \text{either} \ |\Psi_{\beta_f}(t_1)\ra  \ \text{or} \ 0.
% \ \ \text{for} \  \HH_{\mu} \supseteq \HH_\nu.
 \label{be-Pm} \ee
 Substituting Eq.~(\ref{PUP1}) for $\Upsilon_f$ at the time $t_1$ into the right hand side of
 Eq.~(\ref{Psiat1}), then, making use of Eq.~(\ref{be-Pm}), we find
 \be |\Psi_\alpha(t_1)\ra = {\sum_{\beta_f \in g_\alpha(t_1)}} |\Psi_{\beta_f}(t_1)\ra ,
 \label{path-c1}\ee
 where $g_\alpha (t_1)$ is the set of the paths $\beta_f$ for which the components
 $|\Psi_{\beta_f}(t_1)\ra$ lie in the same subspace $\HH_\mu$ as $|\Psi_\alpha(t_1)\ra $
 does, that is,
 \be g_\alpha(t_1) = \{ \beta_f : |\Psi_{\beta_f}(t_1)\ra \in \HH_{\mu},
 \ \text{} \ |\Psi_\alpha(t_1)\ra \in \HH_{\mu} \}. \label{gat}
 \ee

% Beyond the time $t_1$, all the components $|\Psi_\alpha(t)\ra $ and $|\Psi_{\beta_f}(t)\ra $ evolves
% unitarily.
 For times $t$ between $t_1$ and the next splitting times $t_2$ of $\Upsilon$, 
 the tree $\Upsilon_f$ may have some splitting point(s).
 Each of these splitting in $\Upsilon_f$ can be obtained by inserting the identity operator 
 expressed as the summation of a complete set of projection operators [see Eq.(\ref{Psi-alpha})],
 hence, Eq.~(\ref{path-c1}) is still valid for these times $t$, with $g_\alpha(t)$ defined accordingly.
% Furthermore, the principle of compatible description requires that if $|\Psi_\alpha(t)\ra \in \HH_\mu$
% for some R-observable $A_{\{\mu \}}$, then, $|\Psi_{\beta_f}(t)\ra \in \HH_{\mu}$.

 To study the influence of a next splitting point of $\Upsilon$ at $t_2$, we note that
 for times $t\ge t_1$, according to the assumption of specified MsD, the components 
 $|\Psi_\alpha(t)\ra $ in the tree $\Upsilon$ are statistically independent.
 Hence, we can treat each component $|\Psi_\alpha(t)\ra $ of $t \ge t_1$ in the same way as done 
 above for the whole vector $|\Psi(t)\ra $ of $t\ge 0$.
 Then, similarly, we get an equation like Eq.~(\ref{path-c1}) for the time(s) $t_2$,
 with appropriately defined set(s) $g_\alpha(t_2)$.

 Proceeding with the above procedure,
 finally, we get the following result for an arbitrary time $t$:
 For each $\alpha(t) \in \Upsilon$, there exists a corresponding division of the
 set $\{\beta_f(t) \in \Upsilon_f \}$ into subsets $g_{\alpha}(t)$, such that
 \bey \nonumber  |\Psi_\alpha(t)\ra  =  {\sum_{\beta_f \in g_\alpha(t)}} |\Psi_{\beta_f}(t)\ra
 \ \ \hspace{2cm}
 \\  \text{with} \ |\Psi_{\beta_f}(t)\ra \in \HH_{\mu}
 \ \text{if} \ |\Psi_\alpha(t)\ra \in \HH_{\mu} .
 \label{path-c}\eey
% At a splitting time $t_n$, for a component $|\Psi_{\alpha}(t_n)\ra $ given by a projection operator
% $ \PP_{\mu^{\alpha}_{(n)}}$, $|\Psi_{\beta_f}(t_n)\ra \in \HH_{\mu^{\alpha}_{(n)}}$.
 The sets $g_\alpha(t)$ of the paths $\beta_f$, changing at splitting points of $\Upsilon_f$ and $\Upsilon$, 
 have the following properties, 
 \begin{equation}\label{gar}
 \bigcup_{\alpha} g_\alpha(t) = \{ \beta_f(t) \} , \ 
 g_\alpha(t) \bigcap g_{\alpha '}(t) = 0 \ \text{for} \ \alpha \ne \alpha'. 
 \end{equation}
 Equation (\ref{path-c}) shows that $\Upsilon$ can be regarded as a coarse-graining of $\Upsilon_f$.
 Due to the arbitrariness of $\Upsilon$, $\Upsilon_f$ is the finest tree starting from
 the given initial condition.

 Further discussions about the relation between a tree and its coarse-grained trees 
 will be given in the appendix \ref{app-tree}, where it is shown that
 coarse-graining at a splitting point may require further change  in the following part of the tree.

 \subsection{A consequence of the principle of compatible description}

 In this section, we derive a consequence of the principle of compatible description.
 Suppose a tree $\Upsilon$ has a path $\alpha$, at the end of which the system $\R$ has a definite
 value $\mu$ of a R-observable $A_{\{\mu\}}$ at a time $t$.
 It is sometimes useful to indicate explicitly the dependence of the value of $\mu$ on the path
 $\alpha$ and we do this by labeling $\mu_\alpha$. 
 Then,
 \be |\Psi_\alpha(t) \ra \in \HH_{\mu_\alpha} .
% \PP_{\mu_\alpha} |\Psi_\alpha(t)\ra = |\Psi_\alpha(t) \ra.
 \label{PPa} \ee
 For a given value of $\mu$, we use $s_\mu$ to denote the set of paths $\alpha$ for which
 $\mu_\alpha =\mu$, i.e., $s_\mu = \{ \alpha : |\Psi_\alpha(t) \ra \in \HH_{\mu} \}$.
 The component $|\Psi_{\alpha}(t)\ra $ predicts that the system $\R$ has the probability
 [see Eq.(\ref{Pan})]
 \be P_{\Upsilon}(\mu,\alpha,t) = \la \Psi_{\alpha}(t)|\Pm |\Psi_{\alpha}(t)\ra
 \label{P1-oa} \ee
 of possessing the definite value $\mu$.
 The probability $P_\Upsilon(\mu,t)$ for the system $\R$ to have the value $\mu$ at the time $t$ is
 given by $ \sum_{\alpha } P_{\Upsilon}(\mu,\alpha,t)$ and has the following expression,
 \be P_{\Upsilon}(\mu,t) = \sum_{\alpha \in s_\mu} \la \Psi_{\alpha}(t)|\Psi_{\alpha}(t)\ra .
 \label{Pmu-a} \ee

 As discussed in the previous section, there exists a finest specified mixed-state description given
 by the finest tree $\Upsilon_f$.
 Equations (\ref{path-c})-(\ref{PPa}) imply that each component
 $|\Psi_{\beta_f}(t) \ra$ lies in a subspace with a definite value of $\mu$, specifically,
 $|\Psi_{\beta_f}(t) \ra \in \HH_{\mu_\alpha}$ for $\beta_f \in g_\alpha(t)$.
 Let us use $r_\mu$ to denote the set of paths $\beta_f$ whose corresponding components lie in
 the subspace $\HH_\mu$, that is, $r_\mu := \{ \beta_f : |\Psi_{\beta_f}(t) \ra \in \HH_{\mu} \}$.
 Then, the finest tree $\Upsilon_f$ predicts the following probability for the system $\R$
 to have a definite value $\mu$,
 \be P_{\Upsilon_f}(\mu,t) = \sum_{\beta_f \in r_\mu} \la \Psi_{\beta_f}(t)|\Psi_{\beta_f}(t)\ra .
 \label{Pmu-b} \ee
 Making use of the relations given in Eqs.(\ref{path-c}) and (\ref{gar}), it is not difficult to verify
 the following relation between the sets $r_\mu$, $s_\mu$, and $g_\alpha(t)$,
 \be r_\mu = \bigcup_{\alpha \in s_\mu} g_{\alpha}(t). \label{rsg} \ee

 Substituting Eq.~(\ref{path-c}) into Eq.~(\ref{Pmu-a}), we get
 \be P_{\Upsilon}(\mu,t) = \sum_{\alpha \in s_\mu} \ \  \sum_{\beta_f,\beta_f' \in g_{\alpha}(t)}
 \la \Psi_{\beta_f}(t)|\Psi_{\beta_f'}(t)\ra .
 \label{Pmu-abe} \ee
 The principle of compatible description requires that $P_{\Upsilon}(\mu,t)
 = P_{\Upsilon_f}(\mu,t)$.
 Comparing Eq.(\ref{Pmu-b}) with Eq.(\ref{Pmu-abe}) and making use of the relation 
 in Eq.~(\ref{rsg}), we obtain
 \be \sum_{\alpha \in s_\mu} \ \sum_{\beta_f \ne \beta_f' \in g_{\alpha}}
 \la \Psi_{\beta_f}(t)|\Psi_{\beta_f'}(t)\ra \doteq 0 \ \ \forall \mu.
 \label{re-ov} \ee

 Without a complete proof, we conjecture that the validity of Eq.~(\ref{re-ov})
 for all the trees $\Upsilon$ and for all the times $t$ implies the following relation,
 \be \label{D-diag} \D_{\beta_f \beta_f'}  \doteq   \delta_{\beta_f \beta_f'}
 \D_{\beta_f \beta_f}, \ \ \  \ \ \ \forall \beta_f, \beta_f' \in \Upsilon_f,
 \ee
 where
 \bey  \D_{\beta_f \beta_f'} := \la \Psi_{\beta_f}(t)| \Psi_{\beta_f'}(t)\ra .
 \label{Db} \eey
 That is, we conjecture that the principle of compatible description implies Eq.(\ref{D-diag}).
 Note that Eqs.(\ref{path-c}) and (\ref{D-diag}) imply that, for an arbitrary tree $\Upsilon$ with
 paths $\alpha$, 
 \be \label{D-diag-a} \D_{\alpha \alpha'}  \doteq   \delta_{\alpha \alpha'}
 \D_{\alpha \alpha}, \ \ \ \ \forall \alpha , \alpha' \in \Upsilon \ , \ee
 where
 \bey  \D_{\alpha \alpha'} := \la \Psi_{\alpha}(t)| \Psi_{\alpha'}(t)\ra . \label{Da} \eey

 Although we do not have a complete proof for the above conjecture, we do have arguments for
 its correctness.
 In fact, for Eq.~(\ref{re-ov}) to hold for an arbitrary tree $\Upsilon$, generally it is reasonable
 to expect that ${\rm Re} \la \Psi_{\beta_f}(t)|\Psi_{\beta_f'}(t)\ra \doteq 0$ for
 $\beta_f \ne \beta_f'$.
 Then, since $\la \Psi_{\beta_f}(t)|\Psi_{\beta_f'}(t)\ra $ is a dynamical quantity that changes
 with Schr\"{o}dinger evolution except at splitting points, if its real part vanishes for all
 the times, usually its imaginary part should vanish as well.
 The difficulty in completing the proof for Eq.~(\ref{D-diag}) is related to the fact that a mere
 coarse-graining at an arbitrary splitting point of $\Upsilon_f$, without change in other part of the
 tree, does not necessarily give a coarse-grained tree
 (see the appendix \ref{app-tree}), although each $\Upsilon$ is indeed a coarse-graining of $\Upsilon_f$
 for the same initial condition [see Eq.(\ref{path-c})].
 The missing part of the proof is to show that this restriction to the construction of new
 trees by coarse-graining does not influence the validity of Eq.~(\ref{D-diag}).

% \subsection{Mathematical expression of Part I of the third assumption}
 \subsection{Initial-vector restriction}
 \label{sect-cdes}

 In previous sections, making use of the principle of compatible description, we show validity of 
 Eqs.~(\ref{path-c}) and (\ref{D-diag}).
 Now, we show that the principle of compatible description is satisfied, 
 if for trees starting from $|\Psi(t_0)\ra $, 
 there exists a finest tree $\Upsilon_f$ such that both Eqs.~(\ref{path-c}) and (\ref{D-diag}) hold.
 Indeed, when Eq.~(\ref{path-c}) holds, the probability $P_{\Upsilon}(\mu,t)$ is given by
 Eq.~(\ref{Pmu-abe}).
 Then, substituting Eq.~(\ref{D-diag}) into Eq.~(\ref{Pmu-abe}), we have
 \be P_{\Upsilon}(\mu,t) = \sum_{\alpha \in s_\mu} \sum_{\beta_f \in g^\mu_{\alpha}(t)} P_{\beta_f}(t)
 = \sum_{\beta_f \in r_\mu} P_{\beta_f}(t), \label{Pam-b} \ee
 where $P_{\beta_f}(t) = \la \Psi_{\beta_f}(t)|\Psi_{\beta_f}(t)\ra $ is the probability for the
 realization of the path $\beta_f$ in the tree $\Upsilon_f$.
 Therefore, the probability $P_{\Upsilon}(\mu,t)$ can be obtained from related probabilities 
 $P_{\beta_f}(t)$ given by the finest tree, according to the sum rule of probability.
 Making use of this result, it is not difficult to see that all the specified mixed-state descriptions,
 predicted by trees starting from the same initial condition $|\Psi(t_0)\ra $,
 give consistent predictions for the probabilities for $\R$ to have definite values of $\mu$.

 It is straightforward to generalize the above discussions to the case that a fraction of the
 paths of a tree $\Upsilon$ give components possessing definite values of $\mu$.
 The result is similar: The principle of compatible description is satisfied, if
 the two equations (\ref{path-c}) and (\ref{D-diag}) hold.

 Therefore, for a given subsystem $\R$ to be used as the internal measuring apparatus, satisfaction of the 
 principle of compatible description is equivalent to the following restriction to the initial condition.
 \bi
 \item \textbf{Initial-vector restriction}:
 Only those initial vectors are considered, for each of which
 there exists a finest tree $\Upsilon_f$ that satisfies Eq.(\ref{D-diag}) and has the relation
 in Eq.(\ref{path-c}) to all other trees starting from the same initial condition.
 \ei
 Then, the third basic assumption can be expressed as follows: Namely,
 the assumption of specified MsD is applicable to vectors selected by the initial-vector restriction.

 In including this section, we show that the temporary assumption ${\rm A_T}$ given
 in Sec.\ref{sect-R-observable} can be derived from the theory proposed above.
 Let us consider the big system composed of the external observer and $\R+\E$.
 Since there is no interaction between the external observer and $\R+\E$, we assume that
 the big system is initially described by $|\Psi(t_0)\ra \otimes |\xi_0\ra $, where $|\xi_0\ra $ is
 a vector in the Hilbert space of the external observer
 and the vector $|\Psi(t_0)\ra $ satisfies the initial-vector restriction for the system $\R+\E$.
 Suppose application of the assumption of specified MsD predicts paths $|\Psi_\alpha(t)\ra $
 for $\R+\E$, as a result, gives paths $|\Psi_\alpha(t)\ra \otimes |\xi(t)\ra $ for the big system.
% Here, since there is no interaction between the external observer and $\R+\E$,
% there is no correlation between the time evolution of the external observer and of $\R+\E$,
% hence, for the simplicity in discussion we use the same $|\xi(t)\ra $ for all the paths $\alpha$.
 It is not difficult to verify that these paths for the big system also satisfy the
 initial-vector restriction,
 that is, they satisfy both Eqs.(\ref{path-c}) and (\ref{D-diag}),
 therefore, the assumption of specified MsD is applicable to the vector $|\Psi(t)\ra \otimes |\xi(t)\ra $
 for the big system.

 More specifically, if $|\Psi(t)\ra $ satisfies the non-transition condition
 for a time period $\T$, then, the same is true for $|\Psi(t)\ra \otimes |\xi(t)\ra $.
 Therefore, application of the assumption of specified 
 MsD to the big system predicts that the system $\R+\E$ has
 the specified mixed-state description given in Eq.(\ref{rho-ori}).
 This is just what is stated in the temporary assumption ${\rm A_T}$.

 Finally, we give several remarks concerning the initial-vector restriction.
 (1)  In the extreme case that a vector $|\Psi_0\ra $ satisfies the
 initial-vector restriction for none of the subsystems of the total system as the internal measuring 
 apparatus $\R$, there  exists no internal measuring apparatus, hence,
 no possibility of experimentally checking predictions of this vector $|\Psi_0\ra $ from inside.
% In this case, it seems reasonable to assume that such a vector $|\Psi_0\ra $ does not
% describe any physical state of the total system.

 (2) Physical restriction to the initial condition is not a new idea in physics,
 in particular, when dealing with irreversible processes.
 In fact, in studying the microscopic origin of the macroscopic irreversibility stated
 in the second law of
 thermodynamics, it has been suggested by many authors that there might exist some selection rule
 for the initial condition (see, e.g., \cite{Prigogine,Zeh99,BP02}).
% Meanwhile, only for certain types of initial conditions, have master equations been derived
% for open quantum systems \cite{Zeh99,BP02}.
 But, the initial-vector restriction derived in the theory proposed above has not been discussed before. 

 (3) The initial-vector restriction, required by the principle of compatible description, 
 is irrelevant to the restriction used in the definition
 of R-observable for initial product vectors  in Sec.\ref{sect-R-observable}.
 In fact, the latter has nothing to do with the present physical state of the total
 system under consideration.

 \subsection{Further discussions in the condition $C_D$ and the possibility of experimental test}
 \label{sect-NTC}

 In this section, we give further discussions for taking the non-transition condition as  the
 condition $C_D$ in the third basic assumption
 (see Sec.\ref{sect-CD} for previous discussions),
 as well as the possibility of experimental test for this choice of the condition $C_D$.
 It is important to note that the principle of compatible description guarantees the
 physical consistency of the proposed theory, independent of the choice of the condition $C_D$.

 In the theory proposed above, the non-transition condition alone can not guarantee the
 appearance of some definite property of the considered system $\R$.
 In fact, for the system $\R$ to possess a definite property, other (more important) requirements
 must also be met, namely, the principle of compatible description and the existence of R-observable.
 In particular, Eq.(\ref{D-diag}) imposes a restriction
 more stringent than Eq.~(\ref{re-dia}), since the former must be satisfied in all the future times.

 It is experiments that may finally determine whether the non-transition condition is
 sufficient for the condition $C_D$.
 To see this point, let us first consider what may happen, if the non-transition condition
 is looser than what is really needed for the condition $C_D$, but it is still employed 
 as the condition $C_D$.
 In this case, the assumption of specified MsD will
 predict more definite properties than the system $\R$ may really possess.
 In principle, such predictions can be tested experimentally, by studying possible  coherence among
 the components of the specified mixed-state descriptions predicted by the assumption of specified MsD.
 Therefore, at least in principle, experiments may test whether the non-transition condition is
 looser than what is really needed for the condition $C_D$.

 Next, let us discuss what may happen, if some additional requirement is added to the 
 condition $C_D$, in the case that the non-transition condition itself is sufficient for the
 condition $C_D$.
 In this case, the initial-vector restriction will impose more requirements than what are really needed,
 as a result, there will be less vectors that may satisfy the initial-vector restriction.
 The more stringent the condition $C_D$ is, the less there will be valid vectors 
 satisfying the  initial-vector restriction.
 As a result, the predictability and explainablility of the theory will be reduced
 and there may exist experimental results that can not be explained within the theory.
 In fact, in the case of an extremely stringent condition $C_D$, there might exist no vector
 that can satisfy the initial-vector restriction.
 Therefore it is experiments that may finally determine the exact form of the condition $C_D$,
 hence, in principle, the proposed theory is experimentally testable.

 \section{Measurement and measurement results}
 \label{sect-meas-axiom}

 In this section, we discuss measurement processes within the theory proposed above. 
 
 \subsection{General measurement scheme}
 
 A general measurement has the following two basic features:
 (i) In the process of measurement, 
 the interaction between the measuring apparatus $\R$ and the measured system may induce
 transition among subspaces $\HH_\mu$ related to a R-observable $A_{\{\mu\}}$ of $\R$,
 and (ii) after the measurement process, the measuring apparatus has some definite value $\mu$ of 
 the R-observable.

 Let us consider an initial state of the total system described by $|\Psi(t_0)\ra $,
 which satisfies the initial-vector restriction. 
 Suppose the non-transition condition (\ref{NCC}) is satisfied by Schr\"{o}dinger 
 evolution $|\Psi(t)\ra $ within a time interval $[\tau_1,t_1]$
 and with respect to a R-observable $A_{\{ \mu \}}$ of $\R$, 
 where $\tau_1> t_0$ and $t_1-\tau_1>\tau_d$. 
 Then, at the time $t_1$, according to the assumption of specified MsD, 
 the total system $\R+\E$ also has the specified mixed-state description
 that, with a probability $\la \Psi(t_1) |\Pm |\Psi(t_1)\ra$, it is described by 
 $\Pm |\Psi(t_1)\ra $, possessing a definite value $\mu$ of $\R$.
 In principle, from the recorded value $\mu$ of the measuring apparatus $\R$, some information 
 can be obtained about the measured system denote by $\cs$, which is a part of the environment $\E$. 
 We use $\E_1$ to denote the rest part of $\E$, i.e., $\E = \cs + \E_1$.

 To get concrete information from a measurement, there should be further restriction 
 to the interaction process. 
 Of particular interest is a type of interaction process, for which 
 \begin{equation} \label{Pmt1}
 \Pm |\Psi(t_1)\ra = |\phi_\mu (t_1)\ra |\Phi_\mu (t_1)\ra ,
 \end{equation} 
 where $|\phi_\mu(t_1)\ra $ is a vector in $\HH_\cs$, the Hilbert space of the measured system $\cs$, and 
 $|\Phi_\mu (t_1)\ra$ is a normalized vector in $\HH_{\E_1 \R_\mu}$. 
 Here, $\HH_{\E_1 \R_\mu}$ is the direct product of the Hilbert space of the 
 system $\E_1$, denoted by  $\HH_{\E_1}$, and the subspace $\HH_{\R_\mu}$, namely, 
 $\HH_{\E_1 \R_\mu} = \HH_{\E_1} \bigotimes \HH_{\R_\mu}$.
% , where $\HH_{\E_1}$ denotes the Hilbert space of the system $\E_1$. 
 Then, at the time $t_1$,  the measured system $\cs$ lies in a state described 
 by 
 \begin{equation}\label{wpm}
 |\ww\phi_\mu\ra = \frac{|\phi_\mu (t_1)\ra }{ \sqrt{\la \phi_\mu (t_1)|\phi_\mu (t_1)\ra}},
 \end{equation}
 with a probability $p_\mu = \la \phi_\mu (t_1)|\phi_\mu (t_1)\ra $. 
 Thus, with the recorded value $\mu$ of the measuring apparatus $\R$, it is in principle possible
 to infer the state $|\ww\phi_\mu\ra $ of the system $\cs$ after the measurement.

 \subsection{POVM measurement}

 To be able to get, in a more explicit way, information about properties of $\cs$ from the values $\mu$ 
 of the measuring apparatus $\R$, 
 further restriction must be imposed into the measurement scheme. 
 Below, we discuss a measurement scheme, which turns out to be a POVM measurement. 
% For the sake of simplicity in discussion, we consider a R-observable for which 
% $N_\cs = N_A$, where $N_\cs$ is the dimension of the Hilbert space of the measured system $\cs$
% and $N_A$ is the number of the eigenvalues $\mu$ of the considered R-observable of $\R$.

 Let us consider an initial state with a product form, $|\Psi(t_0)\ra = |\phi_0\ra |\Phi_0\ra $, 
 where $|\phi_0\ra \in \HH_\cs$ and $|\Phi_0 \ra \in \HH_{\E_1 \R} \equiv \HH_{\E_1} \bigotimes \HH_{\R}$.
% Suppose the measurement is designed, such that the vectors $|\phi_\mu(t_1)\ra$ are linearly-independent.
% Then, since $N_\cs = N_A$, $|\phi_\mu(t_1)\ra$ form a (unnormalized) basis in $\HH_\cs$, 
% which we denote by $|\phi_m\ra$, with $m=1,\ldots , N_\cs$. 
 Suppose there exist operators $K_\mu$ acting in the Hilbert space $\HH_\cs$, 
 which can connect $|\phi_0\ra $ to $|\phi_\mu(t_1)\ra $, 
 \begin{equation}\label{Km}
 |\phi_\mu(t_1)\ra  = K_{\mu} |\phi_0\ra .
 \end{equation}
 Then, the probability for a state $|\phi_\mu(t_1)\ra $ of $\cs$ to be obtained 
 after a measurement at the time $t_1$ is written as 
 \begin{equation}\label{pmu-K}
 p_\mu = \la \phi_0|K_\mu^\dag K_\mu |\phi_0\ra .
 \end{equation}  
 In the case that the operators $K_\mu$ are independent of the initial vector $|\phi_0\ra $, 
 the unity of the total probability implies that $\sum_\mu K_\mu^\dag K_\mu = I$.
 Thus, we get a POVM measurement, after which with the probability $p_\mu$ in Eq.(\ref{pmu-K})
 a state $K_{\mu} |\phi_0\ra /\sqrt{p_\mu} $ of the system $\cs$ is obtained \cite{nc-book}.
 Furthermore, if the operators $K_\mu$ form a complete set of projection operators in the Hilbert 
 space of the measured system $\cs$, 
 the vectors  $|\ww\phi_\mu\ra $ are orthogonal to each other and the measurement discussed above gives a 
 projective measurement, which is usually discussed in the axiom of measurement in 
 the standard formalism of quantum mechanics.

 Finally, we give some further remarks. 
 (1) The above discussions show that there is no ``collapse of state vector'' in the theory
 proposed above.
 Here, a specified mixed-state description for the total system may appear, because it may
 describe the same physical state of the total system as a pure-vector description $|\Psi(t)\ra $ does.

 (2) Due to the existence of the finest R-observable of the measuring apparatus $\R$, 
 all its R-observables are commutable. 
 Thus, physical observables of the measuring apparatus $\R$ are commutable. 
 This fact is in consistence with our experience that definite properties of a measuring 
 apparatus (within an appropriate energy region) may coexist. 
 On the other hand, for a measured system, discussions given above show that
 the action of measurement is basically similar to that described in the usual quantum mechanics. 
 Hence, as well known, a measured system may have non-commutable observables, which can not have
 definite values at the same time.

 \section{Application I --- Isolatable and non-isolatable systems}
 \label{sect-isolatable}
 \label{sect-NoR}

 In this section, we show that a system which is isolatable from its environment for a sufficiently
 long time period does not have a practically meaningful R-observable.
 We still use $\R$ to denote the considered system in this section, even though a system without
 a R-observable can not be used as a measuring apparatus.

 Consider an initial state described by a normalized vector with a product form,
 $|\Psi(0)\ra = |\psi^{\R}_0\ra |\phi^{\E}_0\ra $.
 We assume that the system $\R$ is isolated from its environment $\E$ within a time period $\T$, 
 which is practically infinitely long, namely, 
 \be \label{ab-nc} H_I|\Psi(t)\ra \doteq 0 \ \ \text{for} \ t \in \T = [0,T]. \ee
 Equation (\ref{ab-nc}) implies that, for $t \in \T$,
 \be e^{-\im H t /\hbar}|\Psi(0)\ra \doteq e^{-\im H_{\R} t /\hbar}|\psi^{\R}_0\ra
 e^{-\im H_{\E} t /\hbar}|\phi^{\E}_0\ra . \label{eH2} \ee
 Let us use $|m_\mu\ra$ to denoted an orthonormal basis in a subspace $\HH_{\R_\mu}$. 
 It is straightforward to verify that elements of the reduced density matrix of $\R$ in these basis states
 can be written as
 \bey \la m_\mu |\rho^{re}_{\R}(t) |n_\nu \ra  = \la \Psi(0)| e^{\im H t /\hbar}|n_\nu \ra
 \la m_\mu |e^{-\im H t /\hbar}|\Psi(0)\ra . \ \
 \label{rhomn} \eey
 Substituting Eq.(\ref{eH2}) into Eq.(\ref{rhomn}), one has
 \bey  \la m_\mu |\rho^{re}_{\R}(t) |n_\nu \ra  \doteq \la \psi^{\R}_0|
 e^{\im H_{\R} t /\hbar}|n_\nu \ra \la m_\mu |e^{-\im H_{\R} t /\hbar}|\psi^{\R}_0\ra , \ \
 \label{rhomn2} \eey
 for $t\in \T$.
 Therefore, usually, Eq.~(\ref{re-dia}) can not hold if the initial vector has
 non-zero components in the two subspaces $\HH_\mu$ and $\HH_\nu$.
 Then, since the time $T$ is practically infinitely long, according to the definition of R-observable,
 at least practically, this system $\R$ does not have a R-observable. 
 Without a R-observable, the assumption of specified MsD is not applicable to the system $\R$,
 hence, such a system can not be used as a measuring apparatus.

 It is straightforward to generalize the above discussions to the case that the time $T$ is long,
 but not practically infinitely long.
 In this case, the system $\R$ may have a R-observable, but with a quite long
 decoherence time $\tau_d$.

 As an example of isolatable system,  let us consider the center-of-mass (COM) degrees of
 freedom of a physical system, with its COM degrees of freedom taken as the system $\R$
 and its internal degrees of freedom taken as a part of the environment of $\R$.
 If it is in principle possible for the COM to be
 uncoupled from its environment for a very long time period such that Eq.~(\ref{ab-nc}) holds,
 then, the COM does not have a R-observable.
 This implies that, in principle, quantum interference effect may be observed for the motion of the
 COM of some appropriately prepared systems, regardless of their masses.
% As long as Eq.~(\ref{ab-nc}) can be satisfied, there is no upper limit for the mass of the system $M$.
 Indeed, up to now, no upper bound has been observed experimentally for the size of a system
 whose COM motion may exhibit quantum interference effects.

 Finally, we discuss briefly some implications of the above results.
 The above discussions show that for a system $\R$ to have a R-observable, it must be
 non-isolatable from its environment.
 In other words, there should exist a part of the environment of $\R$, which is always
 accompanying $\R$ and inducing decoherence to it.
 Some properties that the accompanying environment should have in order to
 guarantee the existence of a R-observable and the validity of Eq.(\ref{D-diag}),
 are discussed in appendixes \ref{sect-decoh-R}, \ref{sect-Daa-fid} and \ref{sect-model}.
 In particular, Eq.(\ref{D-diag}) requires that decoherence effects related a difference in
 some steps of two paths should be able to maintain in all the future times.
 Clearly, this requirement can not be met by a general Hamiltonian,
 hence, it imposes a restriction to the Hamiltonian of the total system.
% As an example, we may consider an atom, which is always interacting with the background
% electromagnetic (radiation) field.
% Indeed, an atom has a quite specific interaction with the radiation field, determined by QED.
% A difficulty met in this study is the ultraviolet divergence;
 %that is, how to treat the renormalization scheme in the study of decoherence.
% This is a quite complex problem and we would leave it to future investigation.

 \section{Application II --- Irreversibility of branching processes}
 \label{sect-irrever}

 In this section, we discuss an effectively-irreversible feature of the proposed theory
 and derive a master equation for the behavior of the measuring apparatus $\R$ in certain ideal processes. 

 \subsection{Irreversible feature of time evolution}
 \label{sect-entropy}

 The branching picture of time evolution discussed in Sec.\ref{sect-branch} is not time-reversible.
 This is not in confliction with the time-reversal symmetry of Schr\"{o}dinger equation,
 because we have the branching picture only for initial conditions selected by the initial-vector 
 restriction with a chosen subsystem as the internal measuring apparatus.

 A quantity that can characterize the irreversible feature of branching
 is von Neumann entropy for the total system,
 \be \label{vN-entropy} S(t) = - {\rm Tr} \{ \rho \ln \rho \}.  \ee
 Substituting the specified mixed-state description $\rho_\Upsilon(t)$ in Eq.~(\ref{rho-fi})
 into Eq.~(\ref{vN-entropy}) and making use of Eq.~(\ref{Pan}) for the probabilities of the realization
 of paths $\alpha$, as well as the orthogonality of paths given in Eq.(\ref{D-diag}), we find
 \be \label{vN-St} S_\Upsilon(t) = - \sum_\alpha P_\alpha(t) \ln P_{\alpha}(t).  \ee
 When no branch-splitting happens, $S_\Upsilon(t)$ keeps constant
 as a result of Schr\"{o}dinger evolution;
 while, at each splitting time $t_i^\alpha$ along a path $\alpha \in \Upsilon$,
 the entropy $S_\Upsilon(t)$ obtains a discontinuous increment.
 Therefore, $S_\Upsilon(t)$ may increase but never decrease with increasing time.
 It is of interest to note that each increment of the entropy $S_\Upsilon(t)$ is related to
 a possibility of measurement.

 The entropy $S_\Upsilon(t)$ in Eq.~(\ref{vN-St}) is different from the thermodynamic entropy.
 (i) $S_\Upsilon(t)$ increases without any upper bound.
 (ii) It is un-measurable; in fact, different branches may predict the same value of $\mu$,
 as a result, when a value $\mu$ comes out as a measurement result, one does not know which path 
 has been realized.

 Coarse-graining may overcome the above-discussed shortcomings.
 For example, consider a time $t$ at which the apparatus $\R$ has definite value of $\mu$.
 Using $p_\mu(t)$ to denote the probability for $\R$ to have a definite value $\mu$ at this time, 
 we have
 \be p_\mu(t) = \sum_{\alpha \ \text{with} \ |\Psi_\alpha(t) \ra \in \HH_\mu} P_\alpha(t) .
 \label{p-mu} \ee
 Using this quantity, we can define an entropy for the measuring apparatus $\R$,
 \be S_{\R}(t) = - \sum_\mu p_\mu(t) \ln p_\mu(t). \label{SR}
 \ee
 This entropy may have an upper bound and is in principle measurable.

 \subsection{Master equation for an ideal case of branching}
 \label{sect-master}

 In this section, we derive a master equation
 for the probabilities for the system $\R$ to take definite values of $\mu$.
 For the simplicity in discussion, we consider an ideal case, in which
 all the time intervals $[\tau_{i+1}^\alpha, \ww \tau_i^\alpha]$ are very short such that
 the non-transition condition is satisfied for almost all the times along the paths,
 as a result, the system $\R$ almost always has definite value of $\mu$. 
 This is a good approximation in some practical situations.
 For the same reason, we assume that the branching times $t_i^\alpha$, as well as the times
 $\tau_i^\alpha$ and $\ww \tau_i^\alpha$ are path-independent, thus,
 we can drop the superscript $\alpha$ in the labeling of these times.
 Moreover, we consider one R-observable $A_{\{\mu\}}$ only.

 Let us consider a path $\alpha$ that ends at a time $t$ beyond a branching time $t_n$,
 with $|\Psi_\alpha(t)\ra $ expressed in Eq.~(\ref{Psi-alpha}).
 At a time $t_{n+1}$, this path splits, resulting in paths we denote by $|\Psi_\beta(t)\ra $
 for $t$ beyond $t_{n+1}$, with
 \bey \beta & = & (\alpha \to \mu_{(n+1)}(t_{n+1})), \
 \\ |\Psi_\beta (t)\ra & = & U(t,t_{n+1}) \PP_{\mu_{(n+1)}} U(t_{n+1},t_n) |\Psi_\alpha (t_n)\ra .
 \label{Psi-b-a}  \eey
 The probability for the realization of a path $\beta$, $P_\beta(t_{n+1})= \la\Psi_\beta (t)
 |\Psi_\beta (t)\ra$, can be written as
 \be P_\beta(t_{n+1}) = \Gamma_n(\alpha, \mu_{(n+1)}) P_\alpha(t_n),
 \label{P-b-a} \ee
 where $\Gamma_n(\alpha, \mu_{(n+1)})$ is defined by
 \bey \nonumber  \Gamma_n(\alpha, \mu) = \hspace{6cm}
 \\  \frac 1{P_\alpha(t_n)}
% \\ & & \ \ \  \cdot
 \la \Psi_\alpha (t_n)|U^\dag (t_{n+1},t_n) \PP_{\mu} U(t_{n+1},t_n)
 |\Psi_\alpha (t_n)\ra .
 \label{Gamma} \eey

 From Eq.~(\ref{p-mu}) we have
 \be p_\mu(t_n) = \sum_{\alpha \ \text{with} \ \mu_{(n)}^\alpha = \mu} P_{\alpha}(t_n).
 \label{pmu} \ee
 For a given value of $\mu_{(n+1)}$, say, $\mu$, the path $\beta $ is determined by
 the path $\alpha $, hence, similar to Eq.~(\ref{pmu}), for the time $t_{n+1}$ we have
 \be p_\mu(t_{n+1}) = \sum_{\alpha} \left . P_{\beta}(t_{n+1}) \right |_{\mu_{(n+1)} = \mu}.
 \label{pmu-be} \ee
 We denote by $\Gamma_n(\mu',\mu)$ the average of $\Gamma_n(\alpha, \mu)$
 over those paths $\alpha $ that have a given value $\mu'$ of $\mu^\alpha_{(n)}$,
 \be \Gamma_n(\mu',\mu) = \frac 1{{\cal N}}
 \sum_{\alpha \ \text{with} \ \mu^{\alpha}_{(n)} = \mu'} \Gamma_n(\alpha, \mu)  ,
 \label{Gamma-mu} \ee
 where
 \be {\cal N} = \sum_{\alpha \ \text{with} \ \mu^{\alpha}_{(n)} = \mu'} 1 . \ee
 Then, we can write
 \be \Gamma_n(\alpha, \mu) = \Gamma_n(\mu^\alpha_{(n)}, \mu)
 + \delta \Gamma_n(\alpha, \mu), \label{gam-n-d} \ee
 where $\delta \Gamma_n(\alpha, \mu)$ denotes deviation of $\Gamma_n(\alpha, \mu)$ from its
 average value $\Gamma_n(\mu^\alpha_{(n)},\mu)$, with average taken
 over paths $\alpha $ having the same value of $\mu^\alpha_{(n)}$.
 Substituting Eq.~(\ref{gam-n-d}) with $\mu =\mu_{(n+1)}$ into Eq.~(\ref{P-b-a}),
 then, substituting the result into Eq.~(\ref{pmu-be}), we have
 \be p_\mu(t_{n+1}) = \sum_\alpha \Gamma_n(\mu^\alpha_{(n)}, \mu) P_\alpha(t_n) + \Delta p ,
 \label{pmu-1} \ee
 where
 \be \Delta p = \sum_{\alpha}  \delta \Gamma_n(\alpha, \mu) P_\alpha(t_n).
 \ee
 The summation over all the paths $\alpha $ is equivalent to a summation over those paths $\alpha $
 with a fixed value of $\mu_{(n)}^\alpha =\mu'$, followed by a summation over $\mu'$.
 Hence, making use of Eq.~(\ref{pmu}), from Eq.~(\ref{pmu-1}) we have
 \be p_\mu(t_{n+1}) = \sum_{\mu'} \Gamma_n(\mu', \mu) p_{\mu'}(t_n) + \Delta p .
 \label{pmu-2} \ee

 To give an estimate to $\Delta p$, we note that
 by definition the average of $\delta \Gamma_n(\alpha, \mu)$ is zero
 and $\sum_\alpha P_\alpha(t) =1$.
 This implies that, when the number of $\alpha $ is sufficiently large,
 $\Delta p$ is usually negligibly small.
 In fact, in the case that $\delta \Gamma_n(\alpha, \mu) P_\alpha(t_n^+)$ can be regarded as
 a random number, one has $\Delta p \sim 1/\sqrt{M_n}$,
 where $M_n$ is the number of the paths $\alpha$.
 It is easy to see that $M_n$ increases exponentially with increasing $n$.

 To summarize, when $n$, the number of steps is sufficiently large, the probability for $\R$ to take
 a definite value of $\mu$ at $t_{n+1}$ satisfies a master equation,
 \be p_\mu(t_{n+1}) \simeq \sum_{\mu'} \Gamma_n(\mu', \mu) p_{\mu'}(t_n) .
 \label{pmu-mas} \ee
 From the definitions given in Eqs.~(\ref{Gamma}) and (\ref{Gamma-mu}),
 it is easy to check that $\Gamma_n(\mu',\mu) \ge 0$ and
 \be \sum_{\mu} \Gamma_n(\mu',\mu) = 1.
 \ee

 Some remarks: Compared with derivations of master equations given in the usual quantum theory,
 the derivation given above has the following advantages:
 It is relatively simple and uses less approximations (we do not need to use approximations like Born
 approximation and Markov approximation).

 \section{Discussions and conclusions}
 \label{sect-conclusion}

 In this section, we first discuss relations between the proposed theory and CHI and MWI of 
 quantum mechanics.
 Then, we give a brief summary for the main results of this paper, as well as some discussions.

 \subsection{Comparison with CHI of quantum mechanics}
 \label{sect-ot-int}

 In this section, we discuss relations between the theory proposed here and CHI.
 In CHI, the time evolution of a quantum system has a stochastic nature and is described by
 (quantum) consistent histories
 \cite{Griffiths84,Grif96,Griff02,Omnes88,Omnes92,Omnes99,GH90,GH93,Hohen09}.
 Each history is composed of a sequence of events represented by time-ordered projection operators,
 with unitary connection between each two successive events.
 The consistency among consistent histories is guaranteed by a consistency condition.
 One projective decomposition of the identity operator, the elements of which are used to 
 construct histories, is called a framework.
 A single-framework rule must be obeyed when CHI is used, which states that a valid description
 must use one framework only, even though other frameworks are also legitimate.

 One may note some similarities between the mathematical formulations of some main results of the theory
 proposed here and of CHI, which we list below.
 \bem
 \item[(1)]
 The description given by a path in the theory here, namely, $|\Psi_\alpha(t)\ra $ in Eq.~(\ref{Psi-alpha}),
 has a formal similarity to the contribution given by a history in CHI.
 \item[(2)] Substituting Eq.~(\ref{Psi-alpha}) into Eq.~(\ref{Da}),
 it is seen that the quantity $\D_{\alpha\alpha'}$ can be written in a form with
 formal similarity to the so-called decoherence functional ${\cal{D}}(\beta , \beta ')$
 in CHI \cite{GH90,GH93},
 \bey \nonumber {\cal{D}}(\beta,\beta') = \tr \left [ P^{(n)}_{\beta_n} U(t_n,t_{n-1})
    \ldots P^{(1)}_{\beta_1} U(t_{1},t_{0}) \rho(t_0)   \right .
 \\  \left . U^{\dag}(t_{1},t_{0}) P^{(1)}_{\beta_1'}
  \cdots  U^{\dag}(t_n,t_{n-1}) P^{(n)}_{\beta_n'}  \right ] , \ \
 \label{Da-ch} \eey
 where $\beta $ indicates a history and $ P^{(j)}_{\beta_j}$
 of $j=1, \ldots , n$ denote projection operators in the history $\beta$.
 The two quantities $\D_{\alpha\alpha}$ and ${\cal{D}}(\beta,\beta)$ give the corresponding
 probabilities, respectively, in the two theories.
 \item[(3)] One of the two requirements of the principle of compatible description, namely,
 Eq.~(\ref{D-diag-a}) has the same formal form as the consistency condition in CHI, which is
 \be \label{CHI-con} {\cal{D}}(\beta , \beta ') =\delta_{\beta \beta'} {\cal{D}}(\beta , \beta ).
 \ee
 \eem

 However, despite the formal similarities mentioned above, the two theories have profound differences
 in their physical contents, as listed below.
% \item[(i)] In the theory here, the principle of compatible description has two requirements, given in
% Eq.(\ref{path-c}) and Eq.(\ref{D-diag}). There is no requirement like Eq.(\ref{path-c}) in CHI.

 (i) {Most of the consistent-histories descriptions allowed in CHI \emph{do not have any} corresponding
 description in the theory proposed here}.
 The reason is as follows.
 In CHI, to have a consistent-histories description of the total system, the only prerequisite
 is given by the consistency condition in Eq.(\ref{CHI-con}) related to some instants (not necessarily
 for all the times).
 It has been pointed out that many of the consistent-histories descriptions do not have a quasiclassical
 feature \cite{DK95} (see also discussions given in Ref.\cite{Grif96}).
 While, the theory proposed here has more stringent restrictions to descriptions of physical states
 of the total system:
 Namely, (a) only R-observables can be used in the assumption of specified MsD to get a specified 
 mixed-state description,
 (b) the non-transition condition must be satisfied around each branching point,
 and (c) the principle of compatible description must be obeyed for all the times.

 For example, in CHI, one is allowed to use a few projection operators related to a few instants
 to construct consistent histories, as long as the histories satisfy the consistency condition in
 Eq.(\ref{CHI-con}) for these instants, in spite of what may happen in future times.
 The consistent-histories descriptions obtained in this way usually do not have any corresponding
 description in the theory proposed here, because
 (a) projection operators used in CHI are not necessarily related to R-observable in the theory here,
 (b) the non-transition condition is not required to be satisfied around the instants considered in CHI,
 (c) more importantly, the descriptions allowed in CHI do not necessarily satisfy the principle
 of compatible description, i.e., Eqs.(\ref{D-diag}) and (\ref{path-c}), in all the future times.

 (ii)  \emph{There is no single-framework rule in the theory proposed here}, while, the
 single-framework rule must be obeyed in CHI to avoid logical inconsistency.
 (There have been some debates in its physical validity \cite{BG-debate,Grif-debate}.)
 In fact, in the theory here, the system $\R$ has a finest R-observable,
 with other R-observables of $\R$ being its coarse-grainings; in the language of CHI,
 this implies that there exists only one legitimate (finest) `framework' for the system $\R$.

 (iii) The initial-vector restriction in the theory here 
 may effectively break the time-reversal symmetry of
 Schr\"{o}dinger equation; while, the time-reversal symmetry is maintained in CHI.

 (iv) In CHI, the consistency condition in Eq.(\ref{CHI-con}) is introduced to guarantee the validity of
 the sum rule of probability.
 In the theory here, the principle of compatible description
 states the physical compatibility of different mathematical descriptions for the same physical
 state of the total system.

 \subsection{Comparison with MWI of quantum mechanics}

% \vspace{0.2cm}
% \noindent \emph{Many-worlds interpretations} ----

 The MWI of quantum mechanics has two main assumptions \cite{Everett57,DG73,Deutsch85}.
 Namely, (i) Schr\"{o}dinger equation holds universally,
 and (ii) the state vector of the total system splits constantly into branches.
 One may combine the MWI and the decoherence theory to get a more complete picture for the time
 evolution, with a branch in MWI related to a preferred (pointer) state (or subspace)
 in the decoherence theory (see, e.g., Ref.\cite{Zurek03}).

 There also exist partial formal similarities between the theory proposed here and the combination
 of MWI and decoherence theory.
 \bi
 \item[(1)] Schr\"{o}dinger equation gives the dynamical law in both theories.
 \item[(2)] By virtue of the assumption of specified MsD,
 the theory here gives a branching picture of time evolution,
 as illustrated in Fig.\ref{fig-bra}, which has a formal similarity to that in MWI.
 \item[(3)] The concept of R-observable has a close relationship to
 the concept of preferred basis (subspace) in the decoherence theory
 \cite{Zurek03,PZ99,GPSS04,Schloss04,WGCL08,SH08,Peres00,JZKGKS03,ZHP93,Eisert04,Zurek81,BHS01},
 though not exactly the same (see App.\ref{sect-RO-PS} for further discussions).
 \ei

 Meanwhile, the two theories have the following main differences:
 \bi
 \item[(i)] In the theory here, the non-transition condition gives an explicitly-expressed
 condition for branching to happen. In MWI, there is no such a condition.
 \item[(ii)] MWI does not have a counterpart of the initial-vector restriction.
 \item[(iv)] In the theory here, there exists only one real world; the description of the world
 may split into branches, but, the real world never splits.
 In MWI, the world may split into many worlds.
 \ei

 Discussions given in this and the previous sections show that the theory proposed here can be
 regarded as certain type of unification of CHI and MWI+decoherence theory \cite{foot-PZ93}.
 However, it is not a direct unification, since the theory here abandons both the single framework
 rule in CHI and the assumption about the splitting of the real world in MWI.
 In addition, the theory here proposes a concrete condition (non-transition condition) for
 the appearance of definite properties, which is given in neither CHI nor MWI.
 Further, unlike in CHI and MWI, in the theory here the internal measuring apparatus must be designated
 and it plays a crucial role in description of the total system.

 \subsection{Summary and discussions}
 \label{sect-sum}

 In this paper, we have proposed a quantum theory for a total system including
 a unique internal measuring apparatus.
 The theory is based on three basic assumptions, which roughly speaking have the following contents:
 (i) the Hilbert space as the state space, (ii) Schr\"{o}dinger equation as the dynamical law,
 and (iii) the assumption of specified MsD for vectors satisfying the initial-vector restriction.

 It has been shown that, when the state of the internal measuring apparatus is concerned, 
 different decomposition of a given density operator into mixture of pure states may have 
 different physical meaning. 
 We use the phase `specified mixed-state description' to call a density operator 
 with a specified decomposition into mixture of pure states. 
 The above mentioned assumption of specified MsD states that some pure-vector descriptions of the total
 system may imply the existence of certain specified mixed-state descriptions.
 The initial-vector restriction is a mathematical expression of the principle of compatible description,
 which states that different mathematical descriptions for the same physical state of the total system
 must give consistent predictions for measurement results of the internal measuring apparatus.

 Loosely speaking, the proposed theory gives the following descriptions for the total system.
 Starting from an initial vector, there always exists a pure
 vector description $|\Psi(t)\ra $, given by Schr\"{o}dinger equation.
 Usually, this pure vector description does not directly give predictions for definite properties
 of the internal measuring apparatus.
 For a physical state of the total system, which is initially described by a
 vector satisfying the initial-vector restriction, the pure-vector description
 $|\Psi(t)\ra $ may imply the existence of certain specified mixed-state descriptions at some times. 
 A specified mixed-state description given in this way may predict some definite property of the 
 internal measuring apparatus, while the pure-vector description $|\Psi(t)\ra $, as a superposition 
 of components in the specified mixed-state description, usually does not.

 The above discussed initial-vector restriction imposes a restriction to initial vectors in the
 Hilbert space for a designated internal measuring apparatus. 
 It may effectively break the time-reversal symmetry of Schr\"{o}dinger equation, leading to the
 irreversibility of some processes, since the time reversal of a vector satisfying the
 initial-vector restriction does not necessarily satisfy the restriction.
 This may shed new light in the old problem of the microscopic origin of the macroscopic
 irreversibility stated in the second law of thermodynamics.

 One characteristic feature of the theory proposed in this paper, distinguishing it from other theories,
 is the role played by the non-transition condition in determining the existence of a
 definite property of the internal measuring apparatus.
 Due to this feature, it is possible that the proposed theory might give some experimentally
 testable predictions. 
% which are not equivalent to the corresponding predictions of other quantum theories.
 However, lots of work are needed before a concrete scheme for experimental test can be proposed, 
 in particular, because of the difficulty in finding solutions of Eq.(\ref{D-diag}).

 \acknowledgments

 The author is grateful to A.J.~Leggett, Yan Gu, Hong Zhao, C.~Kiefer, and Jiangbin Gong
 for valuable discussions and suggestions.
 This work was partially supported by Natural Science Foundation of China
 under Nos.~10775123 and 10975123.

\hspace{1cm}

\appendix

 \section{RSI and CHI do not give concrete condition for definite properties to appear}
 \label{sect-RC-con}

 To be specific, let us discuss a special version of RSI (relative-state interpretations), 
 namely, the many-worlds interpretation (MWI) of quantum mechanics.
 In the theory of MWI, when the state vector for a whole system splits into branches,
 each branch may have a property more definite than that of the whole state vector.
 Thus, in principle, the theory allows the appearance of some definite property of a system.
 However, no concrete condition has been given yet for branching to happen in MWI.
% Without such a condition, MWI can not give the condition for a measuring apparatus to possess
% a definite property.

 It has been suggested that a combination of MWI and the decoherence theory \cite{Zeh70,Zurek81}
 (see reviews given in \cite{JZKGKS03,Zurek03,Schloss04}) may do better,
 with preferred (pointer) states in the decoherence theory related to branches in MWI \cite{Zurek03}.
 However, a concrete condition for branching to happen is still missing, because
 this approach faces the following problem, which is related to the fact that decoherence
 in the decoherence theory is indicated by vanishing
 of off-diagonal elements of the reduced density matrix of a subsystem in certain (preferred) basis.
 That is, a measuring apparatus may be repeatedly used
 and, during each measuring process, some off-diagonal elements of its reduced density matrix
 may become non-negligible due to its interaction with the measured system.
 Therefore, at least for repeatedly used measuring apparatuses, off-diagonal elements of their
 reduced density matrices can not vanish forever.
 Relatedly, the decoherence theory faces a problem stressed by A.J.Leggett, namely,
 distinguishing `false' and `true' decoherence \cite{Leggett02}.

 The theory of CHI considers consistent histories.
 In this theory, when a measuring process is described by a consistent history within a fixed
 framework, a subsystem may have certain definite properties at some times.
 However, starting from a given initial condition,
 there may exist many incompatible frameworks, giving incompatible descriptions (histories).
 To avoid inconsistency, a single-framework rule is assumed, which states that
 one is allowed to adopt only one framework in a consistent discussion,
 though other frameworks are equally valid.
 (The physical origin of this single-framework rule is still not clear and there have been some
 debates in its validity \cite{BG-debate,Grif-debate}.)
 Therefore, the present form of CHI does not supply a method of selecting a specific framework
 from the many incompatible frameworks \cite{KM97}, as a result, it does not give a condition
 under which a definite property of a measuring apparatus may appear.

 \section{A relation between trees and their coarse-grainings}
 \label{app-tree}

 In this appendix, we show that, to obtain a valid coarse-grained tree, in addition to the
 coarse-graining at one step of a path of the original tree, modifications in the subsequent steps
 may also be needed.

 It would be convenient to write explicit dependence of a tree on the initial vector
 $|\Psi(t_0)\ra $ and on the ending time $t$, i.e., write it as $\Upsilon (|\Psi(t_0)\ra ,t)$.
 Let us consider an intermediate time $\ww t_0$, at which the tree $\Upsilon$ gives
 components $|\Psi_\alpha(\ww t_0)\ra $.
 Taking a component $|\Psi_\alpha(\ww t_0)\ra $ as an initial condition,
 the subsequent evolution in the tree $\Upsilon (|\Psi(t_0)\ra ,t)$ forms a sub-tree,
 denoted by $\Upsilon (|\Psi_\alpha(\ww t_0)\ra ,t)$.
 This possibility can be seen clearly in Fig.~\ref{fig-bra}.
 The relation between the original tree and the sub-trees can be written as
 \bey \Upsilon (|\Psi(t_0)\ra,t)  = \bigcup_{\alpha(\ww t_0)} \alpha(\ww t_0) \odot
 \Upsilon (|\Psi_\alpha(\ww t_0)\ra,t), \label{Up-dec}
 \eey
 where we write explicitly the ending time of the paths $\alpha$
 and use $\odot$ to indicate a successive relationship of a path and a following sub-tree.

% Each path $\alpha(\ww t_0)$ extends without splitting, until the next splitting point is met.
 Let us use $B$ to denote the next splitting point of a path $\alpha(\ww t_0)$,
 which takes place at a time
 $t_B$ and is related with a R-observable $A_{\{\mu\}}$; we use $\alpha_B(t)$ of $t\in [\ww t_0,t_B]$
 to indicate the extension of the path $\alpha (\ww t_0)$.
 For $B$ to be a splitting point, the non-transition condition should be satisfied for the R-observable
 $A_{\{\mu\}}$ around the time $t_B$.
 It is easy to see that, around $t_B$,
 the non-transition condition is also satisfied for an arbitrary coarse-grained
 R-observable $A_{\{\eta\}}$ defined using Eq.~(\ref{P-eta}).
 If at the point $B$ we use $A_{\{\eta\}}$, instead of $A_{\{\mu\}}$, to generate the
 splitting of the component $|\Psi_{\alpha_B}( t_B)\ra $, we will get a coarse-grained sub-tree,
 denoted by $\Upsilon_c (|\Psi_{\alpha_B}(\ww t_0)\ra,t)$.
 Then, we get the following coarse-grained tree $\Upsilon_c$ for the initial condition $|\Psi(t_0)\ra$,
 \bey \nonumber
 \Upsilon_c (|\Psi(t_0)\ra,t) =  \alpha_B(\ww t_0) \odot \Upsilon_c (|\Psi_{\alpha_B}(\ww t_0)\ra,t)
 \\  \bigcup_{\alpha \ne \alpha_B} \alpha(\ww t_0) \odot
 \Upsilon (|\Psi_\alpha(\ww t_0)\ra,t), \label{Up-c}
 \eey
 where terms in the second line are the same as the corresponding ones in Eq.~(\ref{Up-dec}).

 Let us compare the sub-tree $\Upsilon (|\Psi_{\alpha_B}(\ww t_0)\ra,t)$ and the coarse-grained
 sub-tree $\Upsilon_c (|\Psi_{\alpha_B}(\ww t_0)\ra,t)$.
 They have the same initial condition $|\Psi_{\alpha_B}(\ww t_0)\ra$ and the same first splitting time
 $t_B$, but they have different R-observables at the splitting time $t_B$, namely,
 $A_{\{\mu\}}$ and  $A_{\{\eta\}}$, respectively.
 Since $A_{\{\eta\}}$ is a coarse-graining of $A_{\{\mu\}}$, the components of the two sub-trees
 at a time $t$ immediately beyond the splitting time $t_B$ have the following relation,
 \be |\Psi_{\ww \alpha}(t)\ra = \Pm |\Psi_{\ww \alpha_c}(t)\ra , \ee
 where we use $\ww \alpha$ and $\ww \alpha_c$ to denote paths in the two sub-trees, respectively.
 Usually, $|\Psi_{\ww \alpha}(t)\ra$ is not equal to $|\Psi_{\ww \alpha_c}(t)\ra$,
 hence, if $|\Psi_{\ww \alpha}(t)\ra$ satisfies the non-transition condition at a time $\ww t_2$
 for some R-observable, it is not necessary for $|\Psi_{\ww \alpha_c}(t)\ra$ to satisfy the
 non-transition condition at the same time and for the same R-observable.
 As a result, the second splitting points of the two sub-trees may be different.
 Therefore, a mere replacement of $\mu(t_B)$ by $\eta(t_B)$ at the point $B$, without any change
 in the following steps, does not necessarily give the coarse-grained tree
 $\Upsilon_c (|\Psi_{\alpha_B}(\ww t_0)\ra,t)$.

 \section{Decoherence mechanism for the existence of R-observable}
 \label{sect-decoh-R}

 As already mentioned previously, Eq.(\ref{re-dia}) in the definition of R-observable
 represents a decoherence effect. 
 In this appendix, we show that this equation 
 can be written in terms of a generalized (quantum) Loschmidt echo.
 Then, we give qualitative arguments for the condition for an observable $A_{\{\mu\}}$ 
 to be a R-observable.

 The left hand side of Eq.~(\ref{re-dia}) can be written in terms of
 quantities in $\HH_{\E}$, by a method similar to that used in a study of
 preferred pointer states \cite{GPSS04,WGCL08}.
 For this purpose, let us consider
 an arbitrary basis in the subspace $\HH_{\R\mu}$, which we denote by $|m_\mu\ra  \in \HH_{\R\mu}$.
 In this basis, Eq.~(\ref{re-dia}) has the following equivalent form,
 \be \la m_\mu | \tr_{\E} \left [ |\Psi (t) \ra \la \Psi (t)| \right ] | n_\nu \ra \doteq 0
 \label{Pmmu} \ee
 for $t\in [\tau_d,T]$ and for all $m_\mu$ and $n_\nu$ with $\mu \ne \nu$.
 Making use of the expression $|\Psi_\mu (t)\ra = \Pm |\Psi(t)\ra $ and the formal solution in
 Eq.(\ref{solu-mu}), it is not difficult to find that Eq.(\ref{Pmmu}) can be written as
 \be \label{PmnP} \la \Psi(0)| e^{ \im H_\nu t/\hbar } |n_\nu\ra
 \la m_\mu | e^{ -\im H_\mu t/\hbar } |\Psi(0)\ra \doteq 0. \ee

 Let us consider initial vectors of the product form, as required in the definition of R-observable,
 \be |\Psi(0)\ra = \left ( \sum_\mu \sum_{m_\mu} c_{m_\mu}(0) |m_\mu\ra \right )
 |\phi_0\ra , \label{ini-prod} \ee
 with $|\phi_0\ra \in \HH_{\E}$.
 If the non-transition condition does not impose too stringent restriction to the coefficients
 $c_{m_\mu}(0)$, then,
 due to the arbitrariness of the coefficients $c_{m_\mu}(0)$, Eq.(\ref{PmnP}) is equivalent to
 the following requirement,
 \be \label{PmnP-2} \la \phi_0| \la n_\nu'| e^{ \im H_\nu t/\hbar } |n_\nu\ra
 \la m_\mu | e^{ -\im H_\mu t/\hbar } |m_\mu'\ra |\phi_0\ra \doteq 0. \ee
 Finally,  introducing the operator
 \be \label{Vmm} V_{m_\mu m_\mu'}(t) := \la m_\mu | e^{ -\im H_\mu t/\hbar } |m_\mu' \ra ,
 \ee
 which represents a non-unitary evolution in the Hilbert space of the environment $\E$,
 we find that Eq.(\ref{re-dia}) has the following form, namely,
 \be \label{gfid-Rb} L_G(t) \doteq 0 \ee
 \text{for}  $t\in [\tau_d,T]$ and for all $m_\mu, m_\mu', n_\nu $ and $n_\nu'$ \text{of}
 $\mu \ne \nu$,  where
 \be L_G(t) := \la \phi_0| V_{n_\nu n_\nu'}^\dag (t) V_{m_\mu m_\mu'}(t) |\phi_0\ra
 \label{MGt} \ee
 is the overlap of two non-unitary evolutions in the Hilbert space of the environment.

 The quantity $L_G(t)$ defined in Eq.~(\ref{MGt}) has a form similar to the so-called quantum
 Loschmidt-echo amplitude.
 To see this point, we recall that,
 as a measure of the stability of the quantum motion of a system under small perturbation,
 the quantum Loschmidt echo is defined as the overlap of the time
 evolution of the same initial state $|\psi(0)\ra $ under two Hamiltonians $H_0$ and
 $H_1$ \cite{Peres84},
 \be M(t) \equiv |m(t)|^2 = |\la \psi(0)|e^{\im H_1t/\hbar} e^{-\im H_0t/\hbar} |\psi(0) \ra |^2.
 \label{Mt} \ee
 Decaying behaviors of the Loschmidt echo with small difference between $H_0$ and $H_1$ have been
 extensively studied in recently years (see review given in Ref.\cite{LE-rev}).
 In particular, when the two systems $H_0$ and $H_1$ are quantum chaotic systems, or quantum
 integrable systems possessing classical counterparts
 with sufficiently large degrees of freedom, the echo has typically an exponential decay
 \cite{fid-chao,CT02,wang,PZ02,LE-rev,WQHW10}.

 Loosely speaking, the following four factors are responsible for the decaying behavior
 of the Loschmidt echo $M(t)$ in Eq.(\ref{Mt}).
 That is, (i) Schr\"{o}dinger evolutions in the two systems $H_0$ and $H_1$ start from the
 same initial state;
 (ii) there exists some difference between $H_0$ and $H_1$;
 (iii) the Hilbert space is sufficiently large, such that there is no finite-dimension
 restriction in the separation of the two trajectories (evolutions) in the Hilbert space.
 In fact, when these three requirements are met, the two systems have different
 trajectories in the Hilbert space, separating with increasing time.
 If, furthermore, the following fourth requirement is met, i.e., (iv) the two systems
 have sufficiently irregular motion in the Hilbert space within the time period of interest,
 then, the Loschmidt echo may have a fast decay, usually, an exponential decay.

 Similar arguments are also applicable to the quantity $L_G(t)$.
 Indeed, $L_G(t)$ is also an overlap of two evolutions starting from the same initial condition.
 The definition in Eq.(\ref{Vmm}) suggests that if the unitary evolution under $H_\mu$
 is sufficiently irregular,
 the operators $V_{m_\mu m_\mu'}(t)$ may generate somewhat irregular motions in the Hilbert space
 of the environment.
 Then, if there are sufficient differences among the effects of $H_\mu$
 and if the Hilbert space of $\E$ is sufficiently large,
 it is reasonable to expect that the quantity $|L_G(t)|$ has a fast decay, like the Loschmidt echo.
 In this case, Eq.~(\ref{gfid-Rb}) may hold, as a result, $A_{\{\mu \}}$ may be a R-observable.

 To summarize, $A_{\{\mu \}}$ may be a R-observable, when the following requirements are met:
 \bem
 \item The operators $V_{m_\mu m_\mu'}(t)$ generate sufficiently irregular motion in the Hilbert space
 of the environment.
 \item The effective Hamiltonians $H_\mu$ have sufficiently different influences in the
 motions generated by $V_{m_\mu m_\mu'}(t)$.
 \item The Hilbert space of the environment $\E$ is sufficiently large.
 \eem

 \section{$\D_{\alpha \alpha'}$ expressed as a generalized Loschmidt echo}
 \label{sect-Daa-fid}

 In this appendix, we give qualitative arguments for that Eq.~(\ref{D-diag-a}) may hold by
 a mechanism similar to that discussed in the previous appendix.
% in particular, $\D_{\alpha \alpha'}$ can be expressed in the form of a generalized Loschmidt echo.
 For this purpose, it would be more convenient to rewrite the component
 $|\Psi_\alpha(t)\ra $ in Eq.~(\ref{Psi-alpha}) in a form with respect to the times $\tau_i$ and
 $\ww\tau_i$.
 The reason is that the non-transition condition (\ref{NCC}) is satisfied within the time intervals
 $[\tau_i^\alpha, \ww\tau_i^\alpha ]$.
 Making use of Eq.~(\ref{solu-mu}) for the time intervals $[\tau_i^\alpha, \ww\tau_i^\alpha ]$,
 we can write $|\Psi_\alpha(t)\ra $ in Eq.~(\ref{Psi-alpha}) in the following form,
 \bey  |\Psi_\alpha(t)\ra = W_\alpha(t,t_0) |\Psi(t_0)\ra  ,
 \label{Pa-exp-U} \eey
 where
 \bey \nonumber W_\alpha(t,t_0) = U(t,\ww \tau_n^\alpha) \
 U_{\mu^\alpha_{(n)}}({\ww \tau_n^\alpha},\tau_n^\alpha)  \
 U(\tau_n^\alpha,{\ww \tau_{n-1}^\alpha})  \hspace{0.8cm}
 \\ \cdots  U_{\mu^\alpha_{(2)}}({\ww \tau_{2}^\alpha},\tau_{2}^\alpha)
 U({\tau_{2}^\alpha},\ww\tau_{1}^\alpha)
 U_{\mu^\alpha_{(1)}}({\ww \tau_{1}^\alpha},\tau_{1}^\alpha) U(\tau_1^\alpha,t_0) , \
 \label{W-al} \label{Pa-exp}
 \eey
 with the following definition of $U_{\mu^\alpha_{(i)}}({\ww \tau_i^\alpha},\tau_i^\alpha)$,
 \be U_{\mu}(t',t)
 :=  \exp \left \{ -\frac{\im}{\hbar}(t'-t)  H_{\mu} \right \}.
 \label{U-mu} \ee
 In the derivation of Eq.(\ref{Pa-exp-U}),  we have used the following property
 related to times $t_i^\alpha \in (\tau_i^\alpha ,\ww\tau_i^\alpha)$, namely,
 \bey  U(\ww\tau_i^\alpha,t_{i}^\alpha) \PP_{\mu^\alpha_{(i)}} U(t_i^\alpha,\tau_i^\alpha)
 =  U_{\mu^\alpha_{(i)}}({\ww \tau_i^\alpha},\tau_i^\alpha),
 \  \eey
 which can be obtained by making use of Eq.~(\ref{Pm-HI}).
 For $t \in (\tau_n^\alpha, \ww \tau_n^\alpha )$, the operator $W_\alpha (t,t_0)$ can be obtained
 by replacing the first two terms on the right hand side of Eq.~(\ref{Pa-exp}) by
 the term $U_{\mu^\alpha_{(n)}}(t,\tau_n^\alpha)$.

 The operator $W_\alpha(t,t_0)$ generates a time evolution in the total Hilbert space,
 given by a sequence of unitary operators $U$ in the total Hilbert space $\HH$, separated by
 unitary operators $U_{\mu}$ acting in subspaces $\HH_{\mu}$.
 As a product of $U$s and $U_\mu$s, $W_\alpha(t,t_0)$ is no longer a unitary operator
 in the total Hilbert space $\HH$.
 Making use of Eq.(\ref{Pa-exp-U}), $D_{\alpha \alpha' }$ can be written as
 \bey D_{\alpha \alpha '} = \la \Psi(t_0)|W^\dag_\alpha(t,t_0) W_{\alpha'}(t,t_0) |\Psi(t_0)\ra  .
 \label{Dab-W} \eey
 Obviously, like $L_G(t)$ discussed in the previous appendix,
 $D_{\alpha \alpha '}$ expressed in Eq.(\ref{Dab-W}) can also be regarded as a generalized
 Loschmidt-echo amplitude, with unitary operators in Eq.(\ref{Mt}) replaced by the operators $W$.
% In the special case of $t \in (\tau_1^\alpha,\ww\tau_1^\alpha)$,
% noticing that all paths of the same tree have the same first splitting point,
%  $\tau^\alpha_1 = \tau^{\alpha'}_1$ and $\mu^\alpha_{(1)} \ne \mu^{\alpha'}_{(1)}$,
% it is easy to verify that $D_{\alpha \alpha'}$ in Eq.~(\ref{Dab-W}) is just a Loschmidt-echo amplitude.

 Then, arguments similar to those given in the previous section for $L_G(t)$ can be applied to the
 quantity $D_{\alpha \alpha '}$, as well, giving the following results:
 Equation (\ref{D-diag-a}) may hold, when the following requirements are met:
 \bem
 \item The operators $W_{\alpha}(t,t_0)$ generate sufficiently irregular motion in the
 total Hilbert space.
 \item The differences among $H_\mu$ should be sufficiently large, such that $W_{\alpha}(t,t_0)$
 of each two different paths generate sufficiently different motions in the Hilbert space.
 \item The total Hilbert space is sufficiently large.
 \eem

% \be U_{\mu^\alpha_{(j)}}({\ww\tau_j^\alpha},\tau_j^\alpha) =
% \exp \left \{ -\frac{i}{\hbar} H_{\mu^\alpha_{(j)}}({\ww\tau_j^\alpha}-\tau_j^\alpha) \right \}
% \ee
% \bw \bey \nonumber |\Psi_\alpha(t)\ra = \exp \left \{ -\frac{i}{\hbar}H(t-{\ww\tau_n^\alpha}) \right \}
% \exp \left \{ -\frac{i}{\hbar} H_{\mu^\alpha_{(n)}}({\ww\tau_n^\alpha}-\tau_n^\alpha) \right \}
% \exp \left \{ -\frac{i}{\hbar}H(\tau_n^\alpha-{\ww\tau_{n-1}^\alpha}) \right \}
% \exp \left \{ -\frac{i}{\hbar} H_{\mu^\alpha_{(n-1)}}({\ww\tau_{n-1}^\alpha}-
% \tau_{n-1}^\alpha) \right \}
% \\ \cdots
% \exp \left \{ -\frac{i}{\hbar} H_{\mu^\alpha_{(1)}}({\ww\tau_1^\alpha}-\tau_1^\alpha) \right \}
% \exp \left \{ -\frac{i}{\hbar}H(\tau_1^\alpha-t_0) \right \} |\Psi(t_0)\ra  . \ \ \
% \label{Pa-exp}
% \eey \ew

 \section{$n$-level system}
 \label{sect-model}

 In this appendix, we discuss a model, in which an $n$-level system interacts with a chaotic environment.
 In this model, more explicit results may be obtained following discussions given in
 Sec.\ref{sect-decoh-R} and Sec.~\ref{sect-Daa-fid}.

 \subsection{Energy eigenstates and R-observable}

 We consider an $n$-level system with normalized energy eigenstates denoted by $|\mu\ra$,
 \be H_\R|\mu\ra =E_\mu|\mu\ra \ \ \ \ \text{for} \ \ \ \mu=1,2,\cdots, n, \ee
 and projection operators $\Pm = |\mu \ra \la \mu |$.
% Clearly, these operators $\Pm$ satisfy the stability condition (\ref{sta}).
 To find out a condition under which the corresponding observable $A_{\{\mu\}}$ can be a
 R-observable defined in Sec.~\ref{sect-R-observable},
 we should consider times $t$ within a time interval, namely, $t\in \T = [0,T]$, within which
 the non-transition condition (\ref{NCC}) is satisfied for an initial vector of a product form,
 $ |\Psi(0)\ra = \left ( \sum_\mu c_\mu |\mu\ra \right ) |\phi(0) \ra \label{Psi0-pro} $,
 where $|\phi(0)\ra $ is a normalized vector in the Hilbert space of the environment $\E$.

 In this case,  off-diagonal elements $\la \mu |\rho^{re}_{\R}|\nu\ra$ with $\mu \ne \nu$
 can be expressed in terms of the Loschmidt echo.
 To show this point, let us write Schr\"{o}dinger evolution $|\Psi(t)\ra $ in the following form,
 \be |\Psi(t)\ra  =  \sum_\mu c_\mu |\mu \ra |\phi_{\mu}(t)\ra . \label{Psi-muphi} \ee
 Substituting Eq.~(\ref{Psi-muphi}) into the definition of the reduced density matrix
 $\rho^{re}_{\R}$, it is ready to find that
 \bey  \langle \mu|\rho^{re}_{\R} (t)|\nu\rangle
  =  c_{\mu } c_{\nu }^* f_{\nu \mu }(t), \label{off-rho} \hspace{0cm} \eey
 where $f_{\nu \mu }(t) =  \la \phi_{\nu }(t)|\phi_{\mu }(t)\ra$.
 To find out an explicit expression for $f_{\nu \mu }(t)$,
 one may substitute Eq.~(\ref{Psi-muphi}) into Eq.~(\ref{SE-Pmu}), getting
 \bey \label{phi-t2} \im \hbar \frac{\pp }{\pp t} |\phi_{\mu}(t)\ra
 \doteq  H_{\mu }^{\E}  |\phi_{\mu }(t)\ra ,
 \eey
 where
 \bey H_{\mu}^{\E} = \la \mu |H|\mu\ra = E_\mu+ H_\E + H^{\E}_{I\mu } \eey
 is a Hermitian operator in the Hilbert space of the environment $\E$, with
 $ H^{\E}_{I\mu} = \la \mu |H_I|\mu\ra \label{HImu} $.
 Hence,
 \be |\phi_\mu(t)\ra  \doteq U_\mu^{\E}(t,0) |\phi(0)\ra , \label{phi-mut} \ee
 where
 \be U_\mu^{\E}(t,0) = \exp \{ -\im t H_\mu^{\E} /\hbar \} . \label{UmE} \ee
 Then, making use of Eq.~(\ref{phi-mut}), it is seen that
 \bey f_{\nu \mu  }(t)  \doteq  \la \phi(0)|e^{\im t H_{\nu }^{\E}
 /\hbar } e^{-\im tH_{\mu }^{\E} / \hbar} |\phi(0)\ra ,
 \label{ft} \eey
 which is a Loschmidt-echo amplitude defined in Eq.~(\ref{Mt}).

 To get an explicit estimate to $|f_{\nu \mu  }(t)|$, let us consider
 an environment that can be modelled by a quantum chaotic system \cite{note-Sred}.
 In this case, the decaying behavior of the Loschmidt echo is separated by
 a perturbative border $\varepsilon_p $, which can be estimated by \cite{CT02}
 \be 2\pi \varepsilon_p \overline{ V_{nd}^2} \sim \sigma_v \Delta , \label{per-bor} \ee
 where $\varepsilon V  = H^{\E}_{I\nu } - H^{\E}_{I\mu } \label{V-define} $
 and $\overline{ V_{nd}^2}$ is the average of $|\la n|V|n'\ra |^2$ with $n \ne n'$.
 Here $|n\ra $ denote the eigenstates of $H_\mu^{\E}$, $\Delta $ is the mean level spacing of
 $H_\mu^{\E}$, and $\sigma_{v}^2$ is the variance of the diagonal elements $\la n|V|n\ra $.
 Below and above the border $\varepsilon_p $, typically, the Loschmidt echo has a Gaussian and
 an exponential decay, respectively,
 \bey |f_{\nu\mu}(t)| \simeq e^{-\varepsilon^2 \sigma^2_{v}t^2/2 \hbar^2},
 \ \ \ \varepsilon < \varepsilon_p , \label{Gaussian}
 \\ |f_{\nu\mu }(t)| \sim e^{-\Gamma t/ 2\hbar }, \ \ \  \ \ \ \varepsilon > \varepsilon_p,
 \label{ft-fgr}
 \eey
 where $\Gamma =2\pi \varepsilon^2 \overline{ V_{nd}^2}/ \Delta $  \cite{fid-chao,CT02,PZ02,wang}.

 For a large environment, $\Delta $ is small, hence, the border $\varepsilon_p$ is low and
 usually one is interested in the case of $\varepsilon > \varepsilon_p$.
 In this case, the Loschmidt echo has the exponential decay in Eq.~(\ref{ft-fgr}),
 as a result, $\langle \mu|\rho^{re}_{\R}(t)|\nu\rangle$ becomes negligibly small for
 times beyond a decoherence time $\tau_d$,
 \bey
 \tau_d = k \hbar \Delta /[\pi \varepsilon^2 \overline{ V_{nd}^2}] ,
 \label{dt-fgr} \eey
 where $k$ is a number determined by the accuracy required.
 This exponential decay of the echo stops when its saturation value is
 reached, which is inversely proportional to the dimension $N$ of the Hilbert space of the environment
 \cite{PZ02}.
 For a sufficiently large environment, the saturation value is (effectively) zero.

 Finally, let us discuss the condition for $A_{\{\mu\}}$ to be a R-observable.
 First, in the case of a constant $V$, $|f_{\nu \mu }(t)| = 1$
 [see Eq.~(\ref{ft})] and $\tau_d = \infty$ [see Eq.~(\ref{dt-fgr})], hence, there is no
 decoherence induced by the environment and Eq.~(\ref{re-dia}) can never be satisfied;
 in this case, $A_{\{\mu\}}$ can not be a R-observable.
 Second, for a non-constant $V$ with a non-zero $\overline{ V_{nd}^2}$,
 the Loschmidt echo decays with time, characterized by the decoherence time
 $\tau_d$ given in Eq.~(\ref{dt-fgr}); in this case, $A_{\{\mu\}}$ can be a R-observable.

 Summarizing the above discussions, we reach the following conclusion:
 For an environment that can be modeled by a quantum chaotic system,
 $A_{\{\mu\}}$ is a R-observable of the system $\R$, if
 the environment is sufficiently large and there are sufficient differences among $H^{\E}_{I\mu }$.
% For a more generic environment with certain irregular property, without proof,
% it seems reasonable to expect that one may have a similar condition for $A_{\{\mu\}}$ to
% be a R-observable of $\R$.

 \subsection{$D_{\beta \alpha}$ expressed in the Hilbert space of the environment}

 The condition under which Eq.(\ref{D-diag-a}) may be satisfied can be discussed in a way similar
 to that given in Appendix \ref{sect-Daa-fid}.
 In the present model, we may express
 $D_{\beta \alpha}$ in terms of quantities in the Hilbert space of the environment.

 For the simplicity in discussion, let us consider an initial vector
 $ |\Psi(t_0)\ra = |\mu_0\ra |\phi_0\ra  \label{Psi0-g} $,
 with $|\phi_0\ra \in \HH_{\E}$.
 Making use of Eq.~(\ref{Pa-exp-U}), the component $|\Psi_\alpha(t)\ra $ can be written as
 \be |\Psi_\alpha(t)\ra = \sum_{\mu} |\mu \ra  |\phi_\mu^\alpha (t)\ra ,
 \ee
 where
 \be |\phi_\mu^\alpha (t)\ra = \la \mu | W_\alpha(t,t_0) |\mu_0 \ra |\phi_0\ra .
 \ee
 Making use of Eq.~(\ref{W-al}), we find that
 \bey \nonumber \la \mu | W_\alpha(t,t_0) |\mu_0 \ra = Y_{\mu \mu^{\alpha}_{(n)}}(t,\ww\tau_n^\alpha)) \
 U^{\E}_{\mu^\alpha_{(n)}}({\ww \tau_n^\alpha},\tau_n^\alpha)  \cdots
% Y_{\mu^{\alpha}_{(n)} \mu^{\alpha}_{(n-1)}}(\tau_n^\alpha,\ww\tau_{n-1}^\alpha)) %
 \\ \cdot  U^{\E}_{\mu^\alpha_{(2)}}({\ww \tau_{2}^\alpha},\tau_{2}^\alpha)
  Y_{\mu^\alpha_{(2)} \mu^\alpha_{(1)}}({\tau_{2}^\alpha},\ww\tau_{1}^\alpha)
 U^{\E}_{\mu^\alpha_{(1)}}({\ww \tau_{1}^\alpha},\tau_{1}^\alpha) \ \ 
 \\ \nonumber \cdot Y_{\mu^\alpha_{(1)} \mu_0 }(\tau_1^\alpha,t_0), \ \ \
 \label{W-al-E}
 \eey
 where $U^{\E}_{\mu}(t,t')$ is defined like the one in Eq.~(\ref{UmE}) and
 \be Y_{\mu \mu'}(t,t') = \la \mu | U(t,t') |\mu' \ra . \label{Ymn} \ee
 The operator $Y_{\mu \mu'}$, though an operator in $\HH_{\E}$, in fact represents transition
 between subspaces $\HH_\mu$ and $\HH_{\mu'}$ in the total Hilbert space.

 Thus, $D_{\beta \alpha}= \la \Psi_\beta(t)|\Psi_\alpha(t)\ra$ has the following expression
 in the Hilbert space of the environment,
 \bey \nonumber D_{\beta \alpha} = \sum_\mu \la \phi_0| Y_{\mu^\beta_{(1)} \mu_0 }^\dag
 U^{\E \dag}_{\mu^\beta_{(1)}} Y_{\mu^\beta_{(2)} \mu^\beta_{(1)}}^\dag
 U^{\E \dag}_{\mu^\beta_{(2)}} \cdots  U^{\E \dag}_{\mu^\beta_{(m)}} Y_{\mu \mu^{\beta}_{(m)}}^\dag
 \\ \cdot Y_{\mu \mu^{\alpha}_{(n)}} U^{\E}_{\mu^\alpha_{(n)}}
 \cdots U^{\E}_{\mu^\alpha_{(2)}} Y_{\mu^\alpha_{(2)} \mu^\alpha_{(1)}} U^{\E}_{\mu^\alpha_{(1)}}
 Y_{\mu^\alpha_{(1)} \mu_0 }|\phi_0\ra , \ \ \ \ 
 \label{Dab-ep} \eey
 where the dependence of $U^{\E}_\mu$ and $Y_{\mu \mu'}$ on times is not written explicitly.

 \section{R-observable and preferred pointer basis}
 \label{sect-RO-PS}

 The basic physical idea behind the concept of R-observable is similar to that behind
 the concept of preferred (pointer) basis in the decoherence theory
 \cite{Zurek03,PZ99,GPSS04,Schloss04,WGCL08,SH08}, more exactly,
 to its generalization as preferred subspace \cite{Peres00}.
% However, it is not an easy task to discuss the difference between the two concepts,
% because there have been so many definitions of preferred basis proposed in the literature.
% Below, we give a brief discussion  for preferred basis defined by
% vanishing of off-diagonal elements of the reduced density matrix in a fixed basis.
 The two concepts have the following main difference:
 The decoherence property of a R-observable given in Eq.(\ref{re-dia})
 is required to hold when the non-transition condition is satisfied.
 In the decoherence theory, the non-transition condition is not a general requirement.

 To be more specific, in the decoherence theory, in the weak coupling limit of the system-environment
 interaction, energy eigenstates of a system with discrete energy levels
 form a preferred basis when some condition is satisfied \cite{PZ99,WGCL08}.
 While, for a quantum Brownian particle with weak coupling with the environment, coherent states
 have been found to be preferred states \cite{Zurek03,JZKGKS03,ZHP93,Eisert04}.
 Furthermore, in the strong coupling limit, eigenstates of the system-environment interaction
 Hamiltonian may form a preferred basis \cite{Zurek81,PZ99,BHS01}.

 Since the non-transition condition generally implies Eq.(\ref{PPHI}),
 the weak coupling limit of the system-environment interaction is a special case of what
 we are considering here.
 Indeed, as shown in Appendix \ref{sect-model}, the finest R-observable of a system $\R$ with $n$ discrete
 energy levels may be related to the energy eigenstates, similar to the case of preferred basis in the
 weak coupling limit in the decoherence theory.
 However, as shown in Sec.\ref{sect-isolatable}, the COM degrees of freedom of a quantum Brownian
 particle does not have a R-observable, in contrast to the case in the decoherence theory with
 coherent states as preferred states.

% This does not conflict with the fact that a measured system may lie in an
% eigenstate of the system-apparatus interaction Hamiltonian after a measurement,
% because R-observable is expected to be a property of the measuring apparatus.
%


\begin{thebibliography}{99}
\bibitem{foot-drb}
 Adopting this starting point, our approach is essentially different 
 from approaches like the dynamical-reduction models, 
 which introduce modifications to Schr\"{o}dinger equation.
 It has an essential difference from the approach of Bohm, too, because in addition to a description given
 in the Hilbert space, the latter requires a hidden variable, i.e., positions of particles.
 (Here we consider only descriptions given in the Hilbert space.)
\bibitem{foot-two-R}
 Relation between descriptions of a total system given by two independent internal observers
 is a more difficult topic.
 In fact, to know the relation, communication between the two observers is necessary.
 Since a measurement should be carried out for a successful communication,
 communications may change the state of the total system.
 This makes the case of two observers much more complex than the case of one observer
 considered here and we would leave this topic for future investigation.
\bibitem{footnote-small-e}
 When assigning definite properties to a subsystem of a total system under Schr\"{o}dinger evolution,
 it is usually unavoidable to face the problem of small probability.
 For example, in the decoherence approach to the foundations of quantum mechanics,
 decoherence may reduce off-diagonal elements of a reduced density
 matrix to small quantities, but can hardly make them exact zero \cite{Schloss04}.
 The theory proposed here belongs to this type of theory.
\bibitem{foot-ips} Obviously, consideration of initial vectors of the product form is not the unique choice.
 But, for the purpose here, i.e., to determine the projection operators $\Pm$ (R-observable),
 this consideration may be enough.
\bibitem{KB07} A different mechanism, by which macrorealism may be achieved, is discussed by 
 J. Kofler and C. Brukner in \prl {\bf 99}, 180403 (2007), where the emphasize is on 
 the influence of limited measurement accuracy in the description of measured systems.
\end{thebibliography}
 \end{document}